# Anatomy-aware and acquisition-agnostic joint registration with SynthMorph

Malte Hoffmann[a,b,c], Andrew Hoopes[a,b,d], Douglas N. Greve[a,b,c], Bruce Fischl[a,b,c,d,*], Adrian V. Dalca[a,b,c,d,*]

[a]Athinoula A. Martinos Center for Biomedical Imaging, Charlestown, MA, United States
[b]Department of Radiology, Massachusetts General Hospital, Boston, MA, United States
[c]Department of Radiology, Harvard Medical School, Boston, MA, United States
[d]Computer Science & Artificial Intelligence Laboratory, Massachusetts Institute of Technology, Cambridge, MA, United States

*These authors contributed equally

Corresponding Author: Malte Hoffmann (mhoffmann@mgh.harvard.edu)

## ABSTRACT

Affine image registration is a cornerstone of medical-image analysis. While classical algorithms can achieve excellent accuracy, they solve a time-consuming optimization for every image pair. Deep-learning (DL) methods learn a function that maps an image pair to an output transform. Evaluating the function is fast, but capturing large transforms can be challenging, and networks tend to struggle if a test-image characteristic shifts from the training domain, such as the resolution. Most affine methods are agnostic to the anatomy the user wishes to align, meaning the registration will be inaccurate if algorithms consider all structures in the image. We address these shortcomings with SynthMorph, a fast, symmetric, diffeomorphic, and easy-to-use DL tool for joint affine-deformable registration of any brain image without preprocessing. First, we leverage a strategy that trains networks with widely varying images synthesized from label maps, yielding robust performance across acquisition specifics unseen at training. Second, we optimize the spatial overlap of select anatomical labels. This enables networks to distinguish anatomy of interest from irrelevant structures, removing the need for preprocessing that excludes content which would impinge on anatomy-specific registration. Third, we combine the affine model with a deformable hypernetwork that lets users choose the optimal deformation-field regularity for their specific data, at registration time, in a fraction of the time required by classical methods. This framework is applicable to learning anatomy-aware, acquisition-agnostic registration of any anatomy with any architecture, as long as label maps are available for training. We analyze how competing architectures learn affine transforms and compare state-of-the-art registration tools across an extremely diverse set of neuroimaging data, aiming to truly capture the behavior of methods in the real world. SynthMorph demonstrates high accuracy and is available at https://w3id.org/synthmorph, as a single complete end-to-end solution for registration of brain magnetic resonance imaging (MRI) data.

## 1. INTRODUCTION

Image registration is an essential component of medical image processing and analysis that estimates a mapping from the space of the anatomy in one image to the space of another image (Cox, 1996; Fischl et al., 2002, 2004; Jenkinson et al., 2012; Tustison et al., 2013). Such transforms generally include an affine component accounting for global orientation such as different head positions, which are typically not of clinical interest. Transforms often include a deformable component that may represent anatomically meaningful differences in geometry (Hajnal & Hill, 2001). Many techniques analyze these further, for example voxel-based morphometry (Ashburner & Friston, 2000; Whitwell, 2009).

Iterative registration has been extensively studied, and the available methods can achieve excellent accuracy both within and across magnetic resonance imaging (MRI) contrasts (Ashburner, 2007; Cox & Jesmanowicz, 1999; Friston et al., 1995; Jiang et al., 1995; Lorenzi et al., 2013; Rohr et al., 2001; Rueckert et al., 1999). Approaches differ in how they measure image similarity and the strategy chosen to optimize it, but the fundamental algorithm is the same: fit a set of parameters modeling the spatial transformation between an image pair by iteratively minimizing a dissimilarity metric. While classical deformable registration can take tens of minutes to several hours, affine registration optimizes only a handful of parameters and is generally faster (Hoffmann et al., 2015; Jenkinson & Smith, 2001; Modat et al., 2014; Reuter et al., 2010). However, these approaches solve an optimization problem for every new image pair, which is inefficient: depending on the algorithm, affine registration of higher-resolution structural MRI, for example, can easily take 5–10 minutes (Table 2). Further, iterative pipelines can be laborious to use. The user typically has to tailor the optimization strategy and choose a similarity metric appropriate for the image appearance (Pustina & Cook, 2017). Often, images require preprocessing, including intensity normalization or removal of structures that the registration should exclude. These shortcomings have motivated work on deep-learning (DL) based registration.

Recent advances in DL have enabled registration with unprecedented efficiency and accuracy (Balakrishnan et al., 2019; Dalca et al., 2018; Eppenhof & Pluim, 2018; Krebs et al., 2017; Li & Fan, 2017; Rohé et al., 2017; Sokooti et al., 2017; Yang et al., 2016, 2017). In contrast to classical approaches, DL models learn a function that maps an input registration pair to an output transform. While evaluating this function on a new pair of images is fast, most existing DL methods focus on the deformable component. Affine registration of the input images is often assumed (Balakrishnan et al., 2019; De Vos et al., 2017) or incorporated *ad hoc*, and thus given less attention than deformable registration (De Vos et al., 2019; Hu et al., 2018; Mok & Chung, 2022; S. Zhao, Dong, et al., 2019; S. Zhao, Lau, et al., 2019). Although state-of-the-art deformable algorithms can compensate for suboptimal affine alignment to some extent, they cannot always fully recover the lost accuracy, as the experiment of Section 4.5 will show.

The learning-based models encompassing both affine and deformable components usually do not consider network generalization to modality variation (De Vos et al., 2019; Shen et al., 2019; S. Zhao, Dong, et al., 2019; S. Zhao, Lau, et al., 2019; Zhu et al., 2021). That is, networks trained on one type of data, such as T1-weighted (T1w) MRI, tend to inaccurately register other types of data, such as T2-weighted (T2w) scans. Even for similar MRI contrast, the domain shift caused by different noise or smoothness levels alone has the potential to reduce accuracy at test time. In contrast, learning frameworks capitalizing on generalization techniques and domain adaptation often do not incorporate the fundamental affine transform (M. Chen et al., 2017; Hoffmann et al., 2022; Iglesias et al., 2013; Qin et al., 2019; Tanner et al., 2018).

A separate challenge for affine registration consists of accurately aligning specific anatomy of interest in the image while ignoring irrelevant content. Any undesired structure that moves independently or deforms non-linearly will reduce the accuracy of the anatomy-specific transform unless an algorithm has the ability to ignore it. For example, neck and tongue tissue can confuse rigid brain registration when it deforms non-rigidly (Andrade et al., 2018; Fein et al., 2006; Fischmeister et al., 2013; Hoffmann et al., 2020).

## 1.1. Contribution

In this work, we present a single, easy-to-use DL tool for fast, symmetric, diffeomorphic—and thus invertible—end-to-end affine and deformable brain registration without preprocessing (Fig. 1). The tool performs robustly across MRI contrasts, intensity scales, and resolutions. We address the domain dependency and anatomical non-specificity of affine registration: while invariance to acquisition specifics will enable networks to generalize to new image types without retraining, our anatomy-specific training strategy alleviates the need for pre-processing segmentation steps that remove image content that would distract most registration methods—as Section 4.4 will show for the example of skull-stripping (Eskildsen et al., 2012; Hoopes, Mora, et al., 2022; Iglesias et al., 2011; Smith, 2002).

Our work builds on ideas from DL-based registration, affine registration, and a recent synthesis-based training strategy that promotes data independence by exposing networks to arbitrary image contrasts (Billot, Greve, et al., 2023; Billot et al., 2020; Hoffmann et al., 2022, 2023; Hoopes, Mora, et al., 2022; Kelley et al., 2024). First, we analyze three fundamental network architectures, to provide insight into how DL models learn and best represent the affine component (Appendix A). Second, we select an optimized affine architecture and train it with *synthetic* data only, making it robust across a landscape of acquired image types without exposing it to any real images during training. Third, we combine the affine model with a deformable hypernetwork to create an end-to-end registration tool, enabling users to choose a regularization strength that is optimal for their own data *without* retraining and in a fraction of the time required by classical methods. Fourth, we test our models across an extremely diverse set of images, aiming to truly capture the variability of real-world data. We compare their performance against popular affine and deformable toolboxes in [Sections 4.4](#) and [4.5](#), respectively, to assess the accuracy users can achieve with off-the-shelf implementations for image types unseen at training.

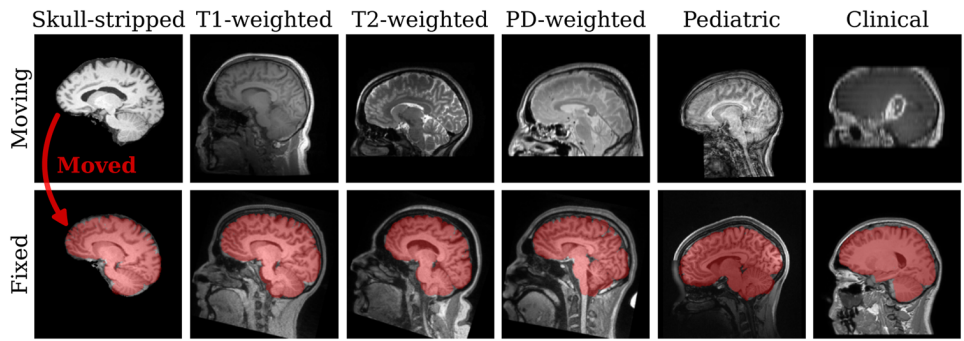

**Fig. 1.** Examples of anatomy-aware SynthMorph affine 3D registration showing the moving brain transformed onto the fixed brain (red overlay). Trained with highly variable synthetic data, SynthMorph generalizes across a diverse array of real-world contrasts, resolutions, and subject populations without any preprocessing.

We freely distribute our source code and tool, Synth-Morph, at [https://w3id.org/synthmorph](https://w3id.org/synthmorph). SynthMorph will ship with the upcoming FreeSurfer release ([Fischl, 2012](#)). For users who wish to use SynthMorph without downloading FreeSurfer, we maintain a standalone container with a wrapper script for easy setup and use supporting any of the following container tools: Docker, Podman, Apptainer, or Singularity.

## 2. RELATED WORK

While this section provides an overview of widely adopted strategies for medical image registration, in-depth review articles are available ([Fu et al., 2020](#); [Oliveira & Tavares, 2014](#); [Wyawahare et al., 2009](#)).

### 2.1. Classical registration

Classical registration is driven by an objective function, which measures similarity in appearance between the moving and the fixed image. A simple and effective choice for images of the same contrast is the mean squared error (MSE). Normalized cross-correlation (NCC) is also widely used, because it provides excellent accu-racy independent of the intensity scale ([Avants et al., 2008](#)). Registration of images across contrasts or modalities generally employs objective functions such as normalized mutual information (NMI) ([Maes et al., 1997](#); [Wells et al., 1996](#)) or correlation ratio ([Roche et al., 1998](#)), as these do not assume similar appearance of the input images. Another class of classical methods uses metrics based on patch similarity ([Glocker et al., 2008, 2011](#); [Ou et al., 2011](#)), which can outperform simpler metrics across modalities ([Hoffmann et al., 2022](#)).

To improve computational efficiency and avoid local minima, many classical techniques perform multi-resolution searches ([Hellier et al., 2001](#); [Nestares & Heeger, 2000](#)). First, this strategy coarsely aligns smoothed downsampled versions of the input images. This initial solution is subsequently refined at higher resolutions until the original images align precisely ([Avants et al., 2011](#); [Modat et al., 2014](#); [Reuter et al., 2010](#)). Additionally, an initial grid search over a set of rotation parameters can help constrain this scale-space approach to a neighborhood around the global optimum ([Jenkinson & Smith, 2001](#); [Jenkinson et al., 2012](#)).

Instead of optimizing image similarity, another registration paradigm detects landmarks and matches these across the images ([Myronenko & Song, 2010](#)). Early work relied on user assistance to identify fiducials ([Besl & McKay, 1992](#); [Meyer et al., 1995](#)). More recent computer-vision approaches automatically extract features ([Machado et al., 2018](#); [Toews & Wells, 2013](#)), for example from entropy ([Wachinger & Navab, 2010, 2012](#)) or difference-of-Gaussians images ([Lowe, 2004](#); [Rister et al., 2017](#); [Wachinger et al., 2018](#)). The performance of this strategy depends on the invariance of landmarks across viewpoints and intensity scales ([Matas et al., 2004](#)).

### 2.2. Deep-learning registration

Analogous to classical registration, unsupervised deformable DL methods fit the parameters of a deep neural network by optimizing a loss function that measures image similarity—but across many image pairs (Balakrishnan et al., 2019; Dalca et al., 2019; De Vos et al., 2019; Guo, 2019; Hoffmann et al., 2022; Krebs et al., 2019). In contrast, supervised DL strategies (Eppenhof & Pluim, 2018; Gopinath et al., 2024; Krebs et al., 2017; Rohé et al., 2017; Sokooti et al., 2017; Yang et al., 2016, 2017) train a network to reproduce ground-truth transforms, for example obtained with classical tools, and tend to underperform relative to their unsupervised counterparts (Hoffmann et al., 2022; Young et al., 2022), although warping features at the end of each U-Net (Ronneberger et al., 2015) level can close the performance gap (Young et al., 2022).

### 2.2.1. Affine deep-learning registration

A straightforward option for an affine-registration network architecture is combining a convolutional encoder with a fully connected (FC) layer to predict the parameters of an affine transform in one shot (Shen et al., 2019; S. Zhao, Dong, et al., 2019; S. Zhao, Lau, et al., 2019; Zhu et al., 2021). A series of convolutional blocks successively halve the image dimension, such that the output of the final convolution has substantially fewer voxels than the input images. This facilitates the use of the FC layer with the desired number of output units, preventing the number of network parameters from becoming intractably large. Networks typically concatenate the input images before passing them through the encoder. To benefit from weight sharing, twin networks pass the fixed and moving images separately and connect their outputs at the end (X. Chen et al., 2021; De Vos et al., 2019).

As affine transforms have a global effect on the image, some architectures replace the locally operating convolutional layers with vision transformers (Dosovitskiy et al., 2020; Mok & Chung, 2022). These models subdivide their inputs into patch embeddings and pass them through the transformer, before a multi-layer perceptron (MLP) outputs a transformation matrix. Multiple such modules in series can successively refine the affine transform if each module applies its output transform to the moving image before passing it onto the next stage (Mok & Chung, 2022). Composition of the transforms from each step produces the final output matrix.

Another affine DL strategy (Moyer et al., 2021; Yu et al., 2021) derives an affine transform without requiring MLP or FC layers, similar to the classical feature extraction and matching approach (Section 2.1). This method separately passes the moving and the fixed image through a convolutional encoder to detect two corresponding sets of feature maps. Computing the barycenter of each feature map yields moving and fixed point clouds, and a least-squares (LS) fit provides a transform aligning them. The approach is robust across large transforms (Yu et al., 2021), while removing the FC layer alleviates the dependency of the architecture on a specific image size.

In this work, we will test these fundamental DL architectures and extend them to build an end-to-end solution for joint affine and deformable registration that is aware of the anatomy of interest.

### 2.3. Robustness and anatomical specificity

Indiscriminate registration of images as a whole can limit the accurate alignment of specific substructures, such as the brain in whole-head MRI. One group of classical methods avoids this problem by down-weighting image regions that cannot be mapped accurately with the chosen transformation model, for example using an iteratively re-weighted LS algorithm (Billings et al., 2015; Gelfand et al., 2005; Modat et al., 2014; Nestares & Heeger, 2000; Puglisi & Battiato, 2011; Reuter et al., 2010). Few approaches focus on specific anatomical features, for example by restricting the registration to regions of an atlas with high prior probability for belonging to a particular tissue class (Fischl et al., 2002). The affine registration tools commonly used in neuroimage analysis (Cox, 1996; Friston et al., 1995; Jenkinson & Smith, 2001; Modat et al., 2014) instead expect—and require—that distracting image content be removed from the input data as a preprocessing step for optimal performance (Eskildsen et al., 2012; Iglesias et al., 2011; Klein et al., 2009; Smith, 2002). Similarly, many DL algorithms assume intensity-normalized and skull-stripped input images (Balakrishnan et al., 2019; Yu et al., 2021; S. Zhao, Lau, et al., 2019), preventing their applicability to diverse unprocessed data.

### 2.4. Domain generalizability

The adaptability of neural networks to out-of-distribution data generally presents a challenge to their deployment (Sun et al., 2016; M. Wang & Deng, 2018). Mitigation strategies include augmenting the variability of the training distribution, for example by adding random noise or applying geometric transforms (Chaitanya et al., 2019; Perez & Wang, 2017; Shorten & Khoshgoftaar, 2019; A. Zhao, Balakrishnan, et al., 2019). Transfer learning adapts a trained network to a new domain by fine-tuning deeper layers on the target distribution (Kamnitsas et al., 2017; Zhuang et al., 2020). These methods require training data from the target domain. By contrast, within medical imaging, a recent strategy synthesizes widely variable training images to promote data independence.

The resulting networks generalize beyond dataset specifics and perform with high accuracy on tasks including segmentation (Billot, Greve, et al., 2023; Billot et al., 2020), deformable registration (Hoffmann et al., 2022), and skull-stripping (Hoopes, Mora, et al., 2022; Kelley et al., 2024). We build on this technology to achieve end-to-end registration.

## 3. METHOD

### 3.1. Background

#### 3.1.1. Affine registration

Let $m$ be a moving and $f$ a fixed image in $N$-dimensional ($N$D) space. We train a deep neural network $h_\theta$ with learnable weights $\theta$ to predict a global transform $T_\theta : \Omega \to \mathbb{R}^N$ that maps the spatial domain $\Omega$ of $f$ onto $m$, given images $\{m, f\}$. The transform $T_\theta = h_\theta(m, f)$ is a matrix

$$T_\theta = \left( \begin{array}{c|c} A_\theta & v_\theta \\ \hline 0 \ \cdots \ 0 & 1 \end{array} \right) = \left( \begin{array}{c|c} t_\theta & \\ \hline 0 \ \cdots \ 0 & 1 \end{array} \right) \quad (1)$$

where matrix $A_\theta \in \mathbb{R}^{N \times N}$ represents rotation, scaling, and shear, and $v_\theta \in \mathbb{R}^{N \times 1}$ is a vector of translational shifts, such that $t_\theta \in \mathbb{R}^{N \times (N+1)}$. We fit the network weights $\theta$ to training set $\mathcal{D}$ subject to

$$\hat{\theta} = \arg\min_\theta \mathbb{E}_{(m,f) \in \mathcal{D}^2} \left[ \mathcal{L}_s\left( m \circ h_\theta(m, f), f \right) \right], \quad (2)$$

where the loss $\mathcal{L}_s$ measures the similarity of two input images, and $m \circ T_\theta$ means $m$ transformed by $T_\theta = h_\theta(m, f)$.

#### 3.1.2. Synthesis-based training

A recent strategy (Billot, Greve, et al., 2023; Billot et al., 2020; Hoffmann et al., 2022, 2023; Hoopes, Mora, et al., 2022) achieves robustness to preprocessing and acquisition specifics by training networks exclusively with synthetic images generated from label maps. From a set of label maps $\{s_m, s_f\}$, a generative model synthesizes corresponding widely variable images $\{m, f\}$ as network inputs. Instead of image similarity, the strategy optimizes spatial label overlap with a (soft) Dice-based loss $\mathcal{L}_o$ (Milletari et al., 2016), strictly independent of image appearance:

$$\mathcal{L}_o\left(\theta, s_m, s_f\right) = -\frac{2}{|J|} \sum_{\substack{j \in J \\ x \in \Omega}} \frac{\left(s_m|_j \circ T_\theta\right)(x) \times s_f|_j(x)}{\left(s_m|_j \circ T_\theta\right)(x) + s_f|_j(x)}, \quad (3)$$

where $s|_j$ represents the one-hot encoded label $j \in J$ of label map $s$ defined at the voxel locations $x \in \Omega$ in the discrete spatial domain $\Omega$ of $f$. The generative model requires only a few label maps to produce a stream of diverse training images that help the network accurately generalize to real medical images of any contrast at test time, which it can register without needing label maps.

### 3.2. Anatomy-aware registration

As we build on our recent work on deformable registration, SynthMorph (Hoffmann et al., 2022), here we only provide a high-level overview and focus on what is new for affine and joint affine-deformable registration. Figure 2 illustrates our setup for affine registration.

#### 3.2.1. Label maps

Every training iteration, we draw a pair of moving and fixed brain segmentations. We apply random spatial transformations to each of them to augment the range of head orientations and anatomical variability in the training set. Specifically, we construct an affine matrix from random translation, rotation, scaling, and shear as detailed in Appendix B.

We compose the affine transform with a randomly sampled and randomly smoothed deformation field (Hoffmann et al., 2022) and apply the composite transform in a single interpolation step. Finally, we simulate acquisitions with a partial field of view (FOV) by randomly cropping the label map, yielding $\{s_m, s_f\}$.

#### 3.2.2. Anatomical specificity

Let $K$ be the complete set of labels in $\{s_m, s_f\}$. To encourage networks to register specific anatomy while ignoring irrelevant image content, we propose to recode $\{s_m, s_f\}$ such that the label maps include only a subset of labels $J \subset K$. For brain-specific registration, $J$ consists of individual brain structures in the deformable case or larger tissue classes in the affine case. At training, the loss $\mathcal{L}$ optimizes only the overlap of $J$, whereas we synthesize images from the complete set of labels $K$, providing rich image content outside the brain as illustrated in Figure 2.

#### 3.2.3. Image synthesis

Given label map $s_m$, we generate image $m$ with random contrast, noise, and artifact corruption (and similarly $f$ from $s_f$). Following SynthMorph, we first sample a mean intensity for each label $j \in K$ in $s_m$ and assign this value to all voxels associated with label $j$. Second, we corrupt $m$ by randomly applying additive Gaussian noise, anisotropic Gaussian blurring, a multiplicative spatial intensity bias field, intensity exponentiation with a global parameter, and downsampling along randomized axes. In aggregate, these steps produce widely varying intensity distributions within each anatomical label (Fig. 3).

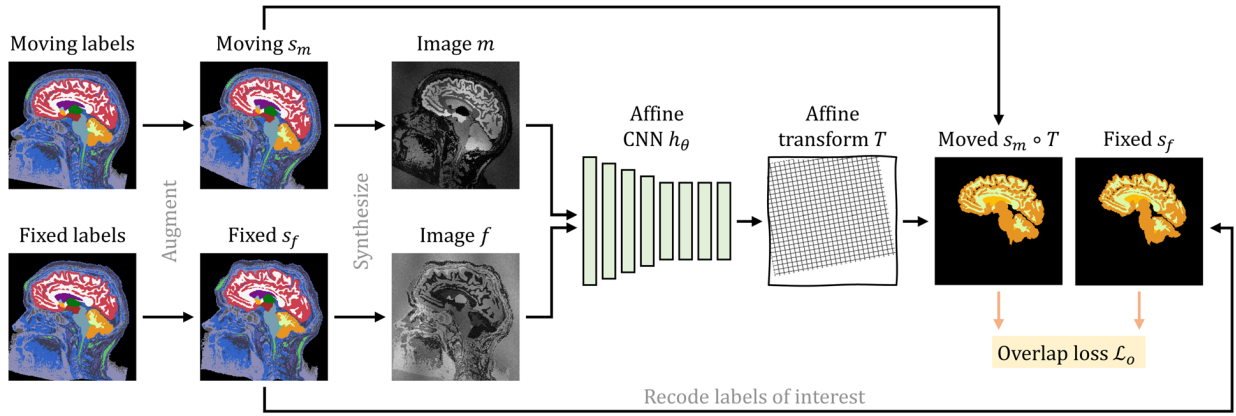

**Fig. 2.** Training strategy for affine registration. At each iteration, we augment a pair of moving and fixed label maps $\{s_m, s_f\}$ and synthesize images $\{m, f\}$ from them. The network $h_\theta$ predicts an affine transform $T$ from which we compute the moved label map $s_m \circ T$ from. Loss $\mathcal{L}_o$ recodes the labels in $\{s_m, s_f\}$ to optimize the overlap of select anatomy of interest only, such as WM, GM, and CSF.

### 3.2.4. Generation hyperparameters

We choose the affine augmentation range such that it encompasses real-world transforms. Appendix Figure A4 (Appendix D) shows the distribution of affine transformation parameters measured across public datasets. We adapt all other values from prior work, which thoroughly analyzed their impact on accuracy (Hoffmann et al., 2022): Appendix Table A2 (Appendix C) lists hyperparameters for label-map augmentation and image synthesis.

### 3.3. Learning

#### 3.3.1. Symmetric affine network

Estimating an affine transform $T$ from a pair of medical images in $ND$ requires reducing a large input space of the order of 100 k–10 M voxels to only $N(N+1)$ output parameters. We extend a recent architecture (Hoffmann et al., 2023; Moyer et al., 2021; A. Q. Wang et al., 2023; Yu et al., 2021), "Detector" in Appendix Figure A1 (Appendix A), that takes a single image as input and predicts a set of $k$ non-negative spatial feature maps $F_i$ with $i \in \{1, 2, ..., k\}$, to support full affine transforms (Yu et al., 2021) *and* weighted least-squares (WLS) (Moyer et al., 2021). Following a series of convolutions, we obtain the center of mass $a_i$ and channel power $p_i |_m$ for each feature map $F_i |_m$ of the moving image,

$$a_i = p_i^{-1} \sum_{x \in \Omega} x F_i |_m (x) \text{ and } p_i |_m = \sum_{x \in \Omega} F_i |_m (x), \quad (4)$$

and separately center of mass $b_i$ with channel power $p_i |_f$ for each $F_i |_f$ of the fixed image. We interpret the sets $\{a_i\}$ and $\{b_i\}$ as corresponding moving and fixed point clouds. Detector refers to a network $h_\theta$ that predicts the affine transform $t_\theta = h_\theta(m, f)$ aligning these point clouds subject to

$$\hat{t}_\theta = \arg \min_t \sum_{i=1}^{k} \epsilon_i \| a_i^\top - (b_i^\top \quad 1) t^\top \|^2, \quad (5)$$

where we use the definition of $t$ from Equation (1) as the submatrix of $T$ that excludes the last row, and we define the normalized scalar weight $\epsilon_i$ as

$$\epsilon_i = p_i |_m \left( \sum_{j=1}^{k} p_j |_m \right)^{-1} p_i |_f \left( \sum_{j=1}^{k} p_j |_f \right)^{-1}. \quad (6)$$

Let $X$ and $y$ be matrices whose $i$th rows are $(a_i^\top \quad 1)$ and $b_i^\top$, respectively. Denoting $W = \text{diag}(\{\epsilon_i\})$, the closed-form WLS solution $\hat{t}_\theta$ of Equation (5) is such that

$$\hat{t}_\theta^\top = (X^\top W X)^{-1} X^\top W y. \quad (7)$$

#### 3.3.2. Symmetric joint registration

For joint registration, we combine the affine model $h_\theta$ with a deformable SynthMorph architecture (Hoffmann et al., 2022). Let $g_\eta$ be a convolutional neural network with parameters $\eta$ that predicts a stationary velocity field (SVF) from concatenated images $\{m, f\}$. While $h_\theta$ predicts symmetric affine transforms by construction, we explicitly symmetrize the SVF:

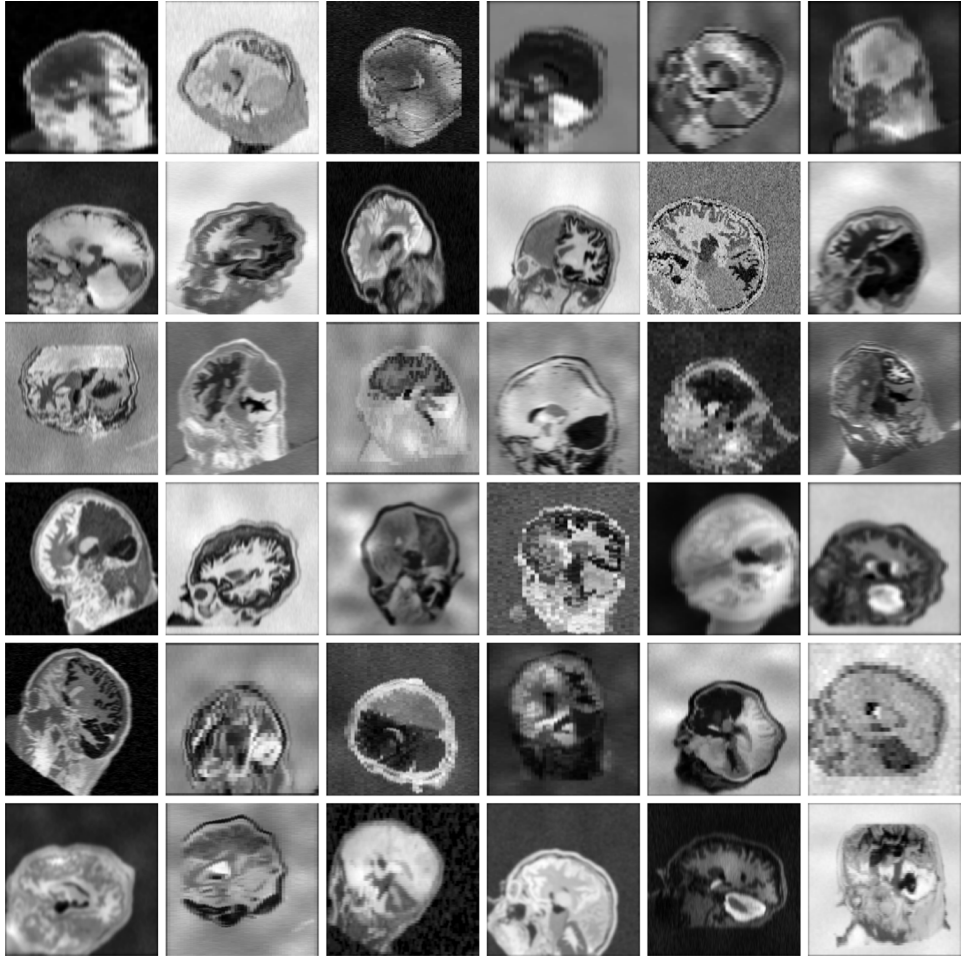

**Fig. 3.** Synthetic 3D training data with arbitrary contrasts, resolutions, and artifact levels, generated from brain label maps. The image characteristics exceed the realistic range to promote network generalization across acquisition protocols. All examples are based on the same label map. In practice, we use several different subjects.

$$\nu_{\eta} = 0.5\Big[g_{\eta}(m,f) - g_{\eta}(f,m)\Big], \tag{8}$$

from which we obtain the diffeomorphic warp field $\phi_{\eta}$ via vector-field integration (Ashburner, 2007; Dalca et al., 2018), and integrating $\nu_{\eta}^{-1} = -\nu_{\eta}$ yields the inverse warp $\phi_{\eta}^{-1}$, up to the numerical precision of the algorithm used. Usually, approaches to learning deformable registration directly fit weights $\eta$ by optimizing a loss of the form

$$\mathcal{L}(\eta, s_m, s_f) = (1-\lambda)\mathcal{L}_o(\phi_{\eta}, s_m, s_f) + \lambda\mathcal{L}_r(\phi_{\eta}), \tag{9}$$

where $\mathcal{L}_o$ quantifies label overlap as before, $\mathcal{L}_r$ is a regularization term that encourages smooth warps, and the parameter $\lambda \in [0, 1]$ controls the weighting of both terms.

Because directly fitting $\eta$ subject to Equation (9) yields an inflexible network predicting warps of fixed regularity, we parameterize $\eta$ using a hypernetwork. Let $\Gamma_{\xi}$ be a neural network with trainable parameters $\xi$. Following our prior work (Hoopes, Hoffmann, et al., 2022; Hoopes et al., 2021), hypernetwork $\Gamma_{\xi}$ takes the regularization weight $\lambda$ as input and outputs the weights $\eta = \Gamma_{\xi}(\lambda)$ of the deformable task network $g_{\eta}$. Consequently, $g_{\eta}$ has no learnable parameters in our setup—its convolutional kernels $\eta$ can flexibly adapt in response to the value $\lambda$ takes at test time.

As shown in Figure 4, for symmetric joint registration, we move images $\{m,f\}$ into an affine mid-space using the matrix square roots of $T_{\theta} = h_{\theta}(m,f)$ and have $g_{\eta}$ predict $\nu_{\eta}$ between images $m \circ T_{\theta}^{-1/2}$ and $f \circ T_{\theta}^{-1/2}$ using kernels $\eta = \Gamma_{\xi}(\lambda)$ specific to input $\lambda$.

While users of SynthMorph can choose between running the deformable step in the affine mid-space or after applying the full transform $T_{\theta}$ to $m$, only the former yields symmetric joint transforms. At training, the total forward transform is $\psi_{\theta\xi} = T_{\theta}^{1/2} \circ \phi_{\xi} \circ T_{\theta}^{1/2}$, and the loss of Equation (9) becomes

$$\mathcal{L}(\lambda, \theta, \xi, \ldots) = (1-\lambda)\mathcal{L}_o(\psi_{\theta\xi}, s_m, s_f) + \lambda\mathcal{L}_r(\phi_{\xi}), \tag{10}$$

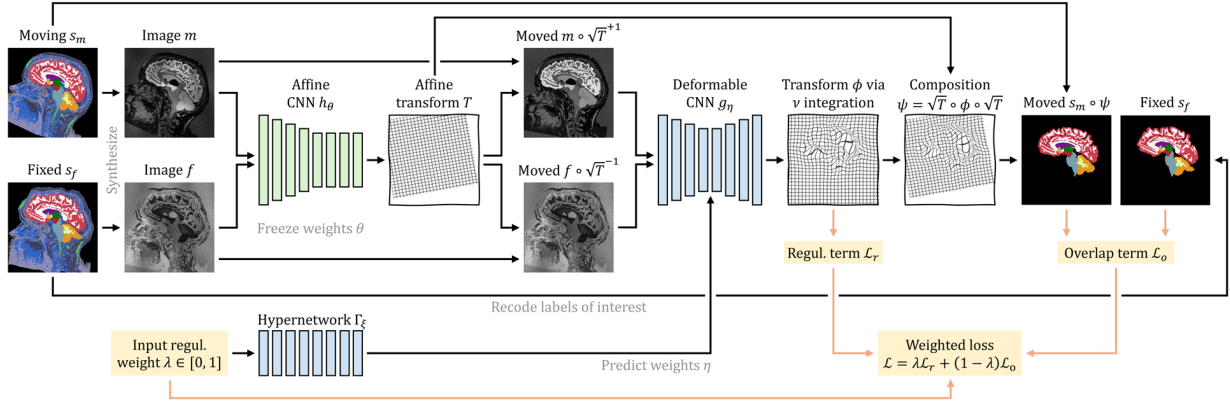

**Fig. 4.** Training strategy for joint registration. As in Figure 2, network $h_\theta$ predicts an affine transform $T$ between moving and fixed images $\{m, f\}$ synthesized from label maps $\{s_m, s_f\}$. Hypernetwork $\Gamma_\xi$ takes the regularization weight $\lambda$ as input and outputs the parameters $\eta = \Gamma_\xi(\lambda)$ of network $g_\eta$. The moved images $m \circ T^{-1/2}$ and $f \circ T^{-1/2}$ are inputs to $g_\eta$, which predicts a diffeomorphic warp field $\phi$. We form the symmetric joint transform $\psi = T^{1/2} \circ \phi \circ T^{-1/2}$ by composition and compute the moved label map $s_m \circ \psi$. Loss $\mathcal{L}_o$ recodes the labels of $\{s_m, s_f\}$ to optimize the overlap of select anatomy of interest—in this case brain labels only.

We choose $\mathcal{L}_r(\phi) = \|\nabla u\|_2$, where $u$ is the displacement of the deformation $\phi = id + u$, and $id$ is the identity field.

### 3.3.3. Overlap loss

In this work, we replace the Dice-based overlap loss term of Equation (3) with a simpler term (Heinrich, 2019; Y. Wang et al., 2021) that measures MSE between one-hot encoded labels $s|_j$,

$$\mathcal{L}_o(\theta, \ldots) = \frac{1}{|\Omega||J|} \sum_{\substack{j \in J \\ x \in \Omega}} \left[ (s_m|_j \circ T_\theta)(x) - s_f|_j(x) \right]^2 \quad (11)$$

where we replace weights $\theta$ with $\{\theta, \xi\}$ and transform $T_\theta$ with $\psi_{\theta\xi}$ for joint registration. MSE is sensitive to the proportionate contribution of each label $j \in J$ to overall alignment, whereas Equation (3) normalizes the contribution of each label by its respective size.

As a result, the MSE loss term discourages the optimization to disproportionately focus on aligning smaller structures, which we find favorable for warp regularity at structure boundaries. In Appendix E, we analyze how optimizing $\mathcal{L}_o$ on label maps compares to an image-similarity loss term.

### 3.3.4. Implementation

Affine SynthMorph implements Detector (Appendix Fig. A1) with $w = 256$ convolutional filters and $k = 64$ output feature maps. The network width $w$ does not vary within the model. We activate the output of each internal block with LeakyReLU (parameter $\alpha = 0.2$) and downsample encoder blocks by a factor of 2 using max pooling.

As in our prior work, the deformable model $g_\eta$ implements a U-Net (Ronneberger et al., 2015) architecture of width $w = 256$, and we integrate the SVF $v_\eta$ via scaling and squaring (Ashburner, 2007; Dalca et al., 2018). Hypermodel $\Gamma_\xi$ is a simple feed-forward network with 4 ReLU-activated hidden FC layers of 32 output units each.

All kernels are of size $3^N$. For computational efficiency, our 3D models linearly downsample the network inputs $\{m, f\}$ and loss inputs $\{s_m, s_f\}$ by a factor of 2. We min-max normalize input images such that their intensities fall in the interval $[0, 1]$. Affine coordinate transforms operate in a zero-centered index space. Appendix B includes further details.

### 3.3.5. Optimization

We fit model parameters with stochastic gradient descent using Adam (Kingma & Ba, 2014) over consecutive training strips $S_i$ ($i \in \{1, 2, \ldots\}$) of $10^6$ batches each. At the beginning of each strip or in the event of divergence, we choose successively smaller learning rates from $l \in \{10^{-4}, 10^{-5}, 10^{-6}\}$. For fast convergence, the first strip of affine training optimizes the overlap of larger label groups than indicated in Section 4.1.3: left hemisphere, right hemisphere, and cerebellum.

Because SynthMorph training is generally not prone to overfitting, it uses a simple stopping criterion measuring progress $P_i$ over batches $t \in S_i$ in terms of validation Dice overlap $D$ (Section 4.3). The 3D models train with a batch size of 1 until the mean overlap across $S_i$ exceeds $P_i = 99.9\%$ of the maximum, that is,

$$P_i = \{|S_i| \max_{t \in S_i} D(t)\}^{-1} \sum_{t \in S_i} D(t). \qquad (12)$$

For joint registration, we uniformly sample hyperparameter values $\lambda \in [0, 1]$. For efficiency, we freeze parameters $\theta$ of the trained affine submodel $h_\theta$ to fit only the weights $\xi$ of hypernetwork $\Gamma_\xi$, optimizing the loss of Equation (10).

However, unfreezing the affine weights within the setup of Figure 4 has no substantial impact on accuracy. Specifically, after one additional strip of joint training, deformable large-21 Dice scores change by $\Delta D \in [-0.5, 0.5]$ depending on the dataset, while affine accuracy decreases by only $\Delta D < 0.1$ points relative to affine-only training.

## 4. EXPERIMENTS

In a first experiment, we train the Detector architecture with synthetic data only. This experiment focuses on building a readily usable tool, and we assess its accuracy in various affine registration tasks. In contrast, Appendix A analyzes the performance of the different architectures across a broad range of variants and transformations, to understand how networks learn and best represent the affine component. In a second experiment, we complete the affine model with a deformable hypernetwork to produce a joint registration solution and compare its performance to readily usable baseline tools.

### 4.1. Data

The training-data synthesis and analyses use 3D brain MRI scans from a broad collection of public data, aiming to truly capture the behavior of the methods facing the diversity of *real-world* images. While users of Synth-Morph do not need to preprocess their data, our experiments use images conformed to the same isotropic $256 \times 256 \times 256$ 1-mm voxel space using trilinear interpolation, and by cropping and zero-padding symmetrically. We rearrange the voxel data to produce gross left-inferior-anterior (LIA) orientation with respect to the volume axes.

#### 4.1.1. Generation label maps

For training-data synthesis, we compose a set of 100 whole-head tissue segmentations, each derived from T1w acquisitions with isotropic ~1-mm resolution. We do not use these T1w images in our experiments. The training segmentations include 30 locally scanned adult FSM subjects (Greve et al., 2021), 30 participants of the cross-sectional Open Access Series of Imaging Studies (OASIS, Marcus et al. 2007), 30 teenagers from the Adolescent Brain Cognitive Development (ABCD) study (Casey et al.,

2018), and 10 infants scanned at Boston Children's Hospital at age 0–18 months (de Macedo Rodrigues et al., 2015; Hoopes, Mora, et al., 2022).

We derive brain label maps from the conformed T1w scans using SynthSeg (Billot, Greve, et al., 2023; Billot et al., 2020). We emphasize that inaccuracies in the segmentations have little impact on our strategy, as the images synthesized from the segmentations will be in perfect voxel-wise registration with the labels by construction.

To facilitate the synthesis of spatially complex image signals outside the brain, we use a simple thresholding procedure to add non-brain labels to each label map. The procedure sorts non-zero image voxels outside the brain into one of six intensity bins, equalizing bin sizes on a per-image basis.

#### 4.1.2. Evaluation images

For baseline comparisons, we pool adult and pediatric T1w images from the Brain Genomics Superstruct Project (GSP, Holmes et al., 2015), the Lifespan Human Connectome Project Development (HCP-D, Harms et al., 2018; Somerville et al., 2018), MASiVar (MASi, Cai et al., 2021) and IXI (Imperial College London, 2015).

The evaluation set also includes IXI scans with T2w and PDw contrast. As all these images are near-isotropic ~1-mm acquisitions, we complement the dataset with contrast-enhanced clinical T1w stacks of axial 6-mm slices from subjects with newly diagnosed glioblastoma (QIN, Clark et al., 2013; Mamonov and Kalpathy-Cramer 2016; Prah et al., 2015).

Our experiments use the held-out test images listed in Table 1. For monitoring and model validation, we use a handful of images pooled from the same datasets, which do not overlap with the test subjects. We do not consider QIN at validation and validate performance in pediatric data with held-out ABCD subjects. To measure registration accuracy, we compute anatomical brain label maps individually for each conformed image volume using SynthSeg (Billot, Greve, et al., 2023; Billot et al., 2020). Although SynthMorph does not require skull-stripping, we skull-strip all images with SynthStrip (Hoopes, Mora, et al., 2022) for a fair comparison across images that have undergone the preprocessing steps expected by the baseline methods—unless explicitly noted.

#### 4.1.3. Labels

The training segmentations encompass a set $K$ of 38 different labels, 32 of which are standard (lateralized) FreeSurfer labels (Fischl et al., 2002). Parenthesizing their average size over FSM subjects relative to the total brain volume and combining the left and right hemispheres, these structures

**Table 1.** Acquired test data for baseline comparisons spanning a range of MRI contrasts, resolutions (res.), and subject populations.

| Dataset | Type | Res. (mm³) | Subjects |
|---------|------|-----------|----------|
| GSP | T1w, age 18–35 a | 1.2×1.2×1.2 | 100 |
| IXI | T1w | 0.9×0.9×1.2 | 100 |
|  | T2w | 0.9×0.9×1.2 | 100 |
|  | PDw | 0.9×0.9×1.2 | 100 |
| HCP-D | T1w, age 5–21 a | 0.8×0.8×0.8 | 80 |
| MASi | T1w, age 5–8 a | 1.0×1.0×1.0 | 80 |
| QIN | post-contrast T1w | 0.4×0.4×6.0 | 50 |

QIN includes contrast-enhanced clinical stacks of thick slices from patients with glioblastoma, whereas the other acquisitions use 3D sequences. While HCP-D and MASi include pediatric data, the remaining datasets consist of adult populations.

are: cerebral cortex (43.4%) and white matter (36.8%), cerebellar cortex (9.2%) and white matter (2.2%), brainstem (1.8%), lateral ventricle (1.7%), thalamus (1.2%), putamen (0.8%), ventral DC (0.6%), hippocampus (0.6%), caudate (0.6%), amygdala (0.3%), pallidum (0.3%), 4th ventricle (0.1%), accumbens (0.1%), inferior lateral ventricle (0.1%), 3rd ventricle (0.1%), and background.

The remaining labels map to variable image features outside the brain (Section 4.1.1). These added labels do not necessarily represent distinct or meaningful anatomical structures but expose the networks to non-brain image content at training. We use all labels $K$ to synthesize training images but optimize the overlap of brain-specific labels $J \subset K$ based on Equation (3).

For affine training and evaluation, we merge brain structures such that $J$ consists of larger tissue classes: left and right cerebral cortex, left and right subcortex, and cerebellum. These classes ensure that small labels like the caudate do not have a disproportionate influence on global brain alignment—different groupings may work equally well. In contrast, deformable registration redefines $J$ to include the 21 largest brain structures up to and including caudate. We use these labels for deformable training and evaluation, as prior analyses report that "only overlap scores of localized anatomical regions reliably distinguish reasonable from inaccurate registrations" (Rohlfing, 2011).

For a less circular assessment of deformable registration accuracy, we separately consider the set of the 10 finest-grained structures above whose overlap we do *not* optimize at training, including the labels from amygdala through 3rd ventricle.

## 4.2. Baselines

We test 3D affine and deformable classical registration with ANTs (Avants et al., 2011) version 2.3.5 using recommended parameters (Pustina & Cook, 2017) for the

NCC metric within and MI across MRI contrasts. We test NiftyReg (Modat et al., 2014) version 1.5.58 with the NMI metric and enable SVF integration for joint registration, as in our approach. We also run the patch-similarity method Deeds (Heinrich et al., 2012), 2022-04-12 version. For a rigorous baseline assessment, we reduce the default grid spacing from 8×7×6×5×4 to 6×5×4×3×2. This setting effectively trades a shorter runtime for increased accuracy as recommended by the author, since it optimizes the parametric B-spline model on a finer control point grid (Heinrich et al., 2013). The modification results in a 1–2% accuracy boost for most datasets as in prior work (Hoffmann et al., 2022). We test affine-only registration with mri_robust_register ("Robust") from FreeSurfer 7.3 (Fischl, 2012) using its robust cost functions (Reuter et al., 2010), as only the robust cost functions can down-weight the contribution of regions that deform non-linearly. However, we highlight that the robust-entropy metric for cross-modal registration is experimental. We use Robust with up to 100 iterations and initialize the affine registration with a rigid run. Finally, we also test affine and deformable registration with the FSL (Jenkinson et al., 2012) tools FLIRT (Jenkinson & Smith, 2001) version 6.0 and FNIRT (Andersson et al., 2007) build 507. While the recommended cost function of FLIRT, correlation ratio, is suitable within and across modalities, we emphasize that users cannot change FNIRT's MSE objective, which specifically targets within-contrast registration.

We compare DL model variants covering popular registration architectures in Section A.3. This analysis uses the same capacity and training set for each model. For our final synthesis-based tool in Sections 4.4 and 4.5, we consider readily available machine-learning baselines trained by their respective authors, to assess their generalization capabilities to the diverse data we have gathered. This strategy evaluates what level of accuracy a user can expect from off-the-shelf methods without retraining, as retraining is generally challenging for users (see Section 5.3). We test: KeyMorph (Yu et al., 2021) and C2FViT (Mok & Chung, 2022) models trained for pair-wise affine, and the 10-cascade Volume Tweening Network (VTN) (S. Zhao, Dong, et al., 2019; S. Zhao, Lau, et al., 2019) trained for joint affine-deformable registration. Each network receives inputs with the expected image orientation, resolution, and intensity normalization.

In contrast to the baselines, SynthMorph is the only method optimizing spatial label overlap. While this likely provides an advantage when measuring accuracy with a label-based metric, optimizing an image-based objective may be advantageous when measuring image similarity at

**Table 2.** Single-threaded runtimes on a 2.2-GHz Intel Xeon Silver 4114 CPU, averaged over $n = 10$ runs.

| Method | Affine (seconds) | Deformable (seconds) |
|---|---|---|
| ANTs | $777.8 \pm 36.0$ | $17189.5 \pm 52.7$ |
| NiftyReg | $293.7 \pm 0.5$ | $7021.0 \pm 21.3$ |
| Deeds | $142.8 \pm 0.3$ | $383.1 \pm 0.6$ |
| Robust | $1598.9 \pm 0.8$ | – |
| FSL | $151.7 \pm 0.4$ | $8141.5 \pm 195.7$ |
| C2FViT[a] | $43.7 \pm 0.3$ | – |
| KeyMorph | $36.2 \pm 2.6$ | – |
| VTN[b] | – | $63.5 \pm 0.3$ |
| SynthMorph | $72.4 \pm 0.8$ | $887.4 \pm 2.5$ |

Errors indicate standard deviations. On an NVIDIA V100 GPU, all affine and deformable DL runtimes (bottom) are ~1 minute, including setup.
[a]Timed on the GPU as the device is hard-coded.
[b]Implementation performs joint registration only.

test. For a balanced comparison, we assess registration accuracy in terms of label overlap and image similarity.

### 4.3. Evaluation metrics

To measure registration accuracy, we propagate the moving label map $s_m$ using the predicted transform $T$ to obtain the moved label map $s_m \circ T$ and compute its (hard) Dice overlap $D$ (Dice, 1945) with the fixed label map $s_f$. In addition, we evaluate MSE of the modality-independent neighborhood descriptor (MIND, Heinrich et al., 2012) between the moved image $m \circ T$ and the fixed image $f$ as well as NCC for same-contrast registration. As we seek to measure brain-specific registration accuracy, we remove any image content external to the brain labels prior to evaluating the image-based metrics. We use paired two-sided $t$-tests to determine whether differences in mean scores between methods are significant.

We analyze the regularity of deformation field $\phi$ in terms of the mean absolute value of the logarithm of the Jacobian determinant $J_\phi$ over brain voxels $\Omega_B$. This quantity is sensitive to the deviation of $J_\phi$ from the ideal value 1 and thus measures the width of the distribution of log-Jacobian determinants, the "log-Jacobian spread" $\delta$:

$$\delta(\phi) = \frac{1}{|\Omega^*|} \sum_{x \in \Omega_B^*} |\ln(|J_\phi(x)|)|, \tag{13}$$

where $\Omega_B^* = \{x \in \Omega_B \mid J_\phi(x) \neq 0\}$. We also determine the proportion of folding voxels, that is, locations $x \in \Omega_B$ where $J_\phi(x) < 0$. We compare the inverse consistency of registration methods by means of the average displacement $E$ that voxels undergo upon subsequent application of transforms $\{T_1, T_2\}$,

$$E(T_1, T_2) = \frac{1}{|\Omega_B|} \sum_{x \in \Omega_B} \| (T_2 \circ T_1)(x) - x \|_2. \tag{14}$$

Specifically, we evaluate the mean symmetric inverse consistency $I$ of method $h$ with $T_1 = h(m, f)$ and $T_2 = h(f, m)$ for any pair of input images $\{m, f\}$:

$$I(h, m, f) = 0.5 \left[ E(T_1, T_2) + E(T_2, T_1) \right]. \tag{15}$$

### 4.4. Experiment 1: affine registration

In this experiment, we focus on "affine SynthMorph," an anatomy-aware affine registration tool that generalizes across acquisition protocols while enabling brain registration without preprocessing. In contrast, Appendix A compares competing network architectures and analyzes how they learn and best represent affine transforms.

#### 4.4.1. Setup

First, to give the reader an idea of the accuracy achievable with off-the-shelf algorithms for data unseen at training, we compare affine SynthMorph to classical and DL baselines trained by the respective authors. We test affine registration of skull-stripped images across MRI contrasts, for a variety of different imaging resolutions and populations, including adults, children, and patients with glioblastoma. We also compare the symmetry of each method with regard to reversing the order of input images. Each test involves held-out image pairs from separate subjects, summarized in Table 1.

Second, we analyze the effect of thick-slice acquisitions on affine SynthMorph accuracy compared to classical baselines. This experiment retrospectively reduces the through-plane resolution of the moving image of each GSP→IXI$_{T_1}$ pair to produce stacks of axial slices of thickness $\Delta z \in \{1, 2, ..., 10\}$ mm. At each $\Delta z$, we simulate partial voluming (Kneeland et al., 1986; Simmons et al., 1994) by smoothing all moving images in slice-normal direction with a 1D Gaussian kernel of full-width at half-maximum (FWHM) $\Delta z$ and by extracting slices $\Delta z$ apart using linear interpolation. Finally, we restore the initial volume size by linearly upsampling through-plane.

Third, we evaluate the importance of skull-stripping the input images for accurate registration. With the exception of skull-stripping, we preprocess full-head GSP→IXI$_{T_1}$ pairs as expected by each method and assess brain-specific registration accuracy by evaluating image-based metrics within the brain only.

#### 4.4.2. Results

Figure 6 shows representative registration examples for the tested dataset combinations, while Figure 5 quantitatively compares affine registration accuracy across skull-stripped image pairings. Although affine SynthMorph has not seen any real MRI data at training, it achieves the highest Dice score for every dataset tested.

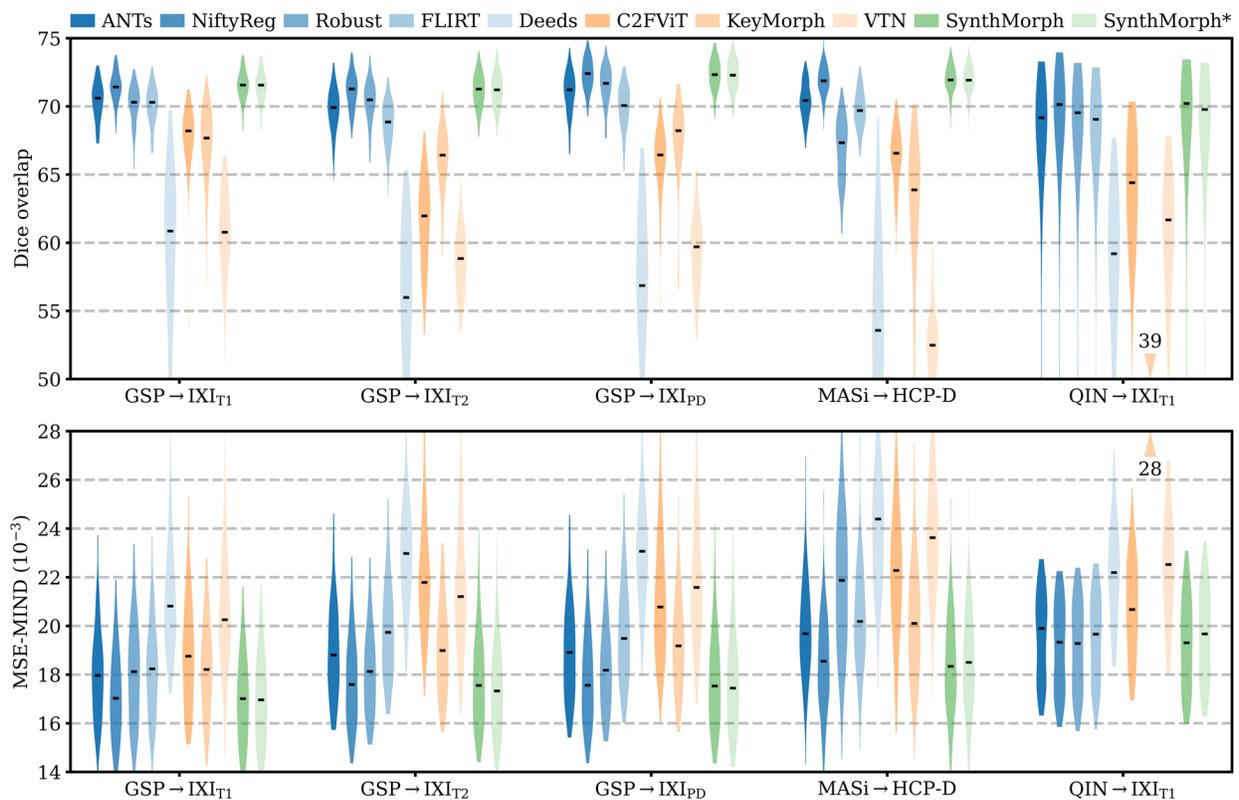

**Fig. 5.** Affine 3D registration accuracy as mean Dice scores and in terms of image similarity. Each violin shows the distribution across the skull-stripped cross-subject pairs from Table 1. For comparison, the asterisk indicates SynthMorph performance without skull-stripping. Downward arrows indicate median scores outside the plotted range. Higher Dice and lower MSE-MIND are better.

For the GSP→IXI$_{T1}$ and MASi→HCP-D pairs that most baselines are optimized for, SynthMorph exceeds the best-performing baseline NiftyReg by $\Delta D = 0.1$ points ($p < 10^{-4}$ and $p < 0.03$ for paired two-sided $t$-tests). Across all other pairings, SynthMorph matches the Dice score achieved by the most accurate affine baseline, which is NiftyReg in every case. Method Deeds performs least accurately, lagging behind the second last classical baselines by $\Delta D = 10.1$ or more. The other classical methods perform robustly across all testsets, generally within 1–2 Dice points of each other.

On the MASi→HCP-D testset, FLIRT's performance exceeds Robust by $\Delta D = 3$ ($p = 10^{-23}$) and matches it across GSP→IXI$_{T1}$ pairs ($p = 0.8$). Across the remaining testsets, FLIRT ranks fourth among classical baselines.

In contrast, the DL baselines do not reach the same accuracy. Even for the T1w pairs they were trained with, SynthMorph leads by $\Delta D = 3.7$ or more, likely due to domain shift between the test and baseline training data. As expected, DL-baseline performance continues to decrease as the test-image characteristics deviate further from those at training. Interestingly, VTN consistently ranks among the least accurate baselines, although its preprocessing effectively initializes the translation and scaling parameters by separately adjusting the moving and fixed images such that the brain fills the whole FOV.

Even though affine SynthMorph does not directly optimize image similarity at training, it surpasses NiftyReg for GSP→IXI$_{T1}$ ($p < 2×10^{-11}$) and MASi→HCP-D ($p < 0.02$) pairs in terms of the image-based MSE-MIND metric. Generally, MSE-MIND ranks the methods similarly to Dice overlap, as does NCC across the T1w registration pairs (Fig. 8a).

Figure 9 shows that SynthMorph's affine transforms across GSP→IXI$_{T1}$ are more symmetric than all baselines tested. When we reverse the order of the input images, the mean inconsistency between forward and backward transforms is $I = 5×10^{-5}$ mm per brain voxel, closely followed by NiftyReg. Robust also uses an inverse-consistent algorithm, leading to $I = 8×10^{-3}$ mm. The remaining baselines are substantially less symmetric, with inconsistencies of $I = 2$ mm for KeyMorph or more.

Figure 7a shows how registration accuracy evolves with increasing moving-image slice thickness $\Delta z$. SynthMorph and ANTs remain the most robust for $\Delta z \leq 6$ mm, reducing only to 99% at $\Delta z = 10$ mm. For $\Delta z \in [2, 5]$ mm,

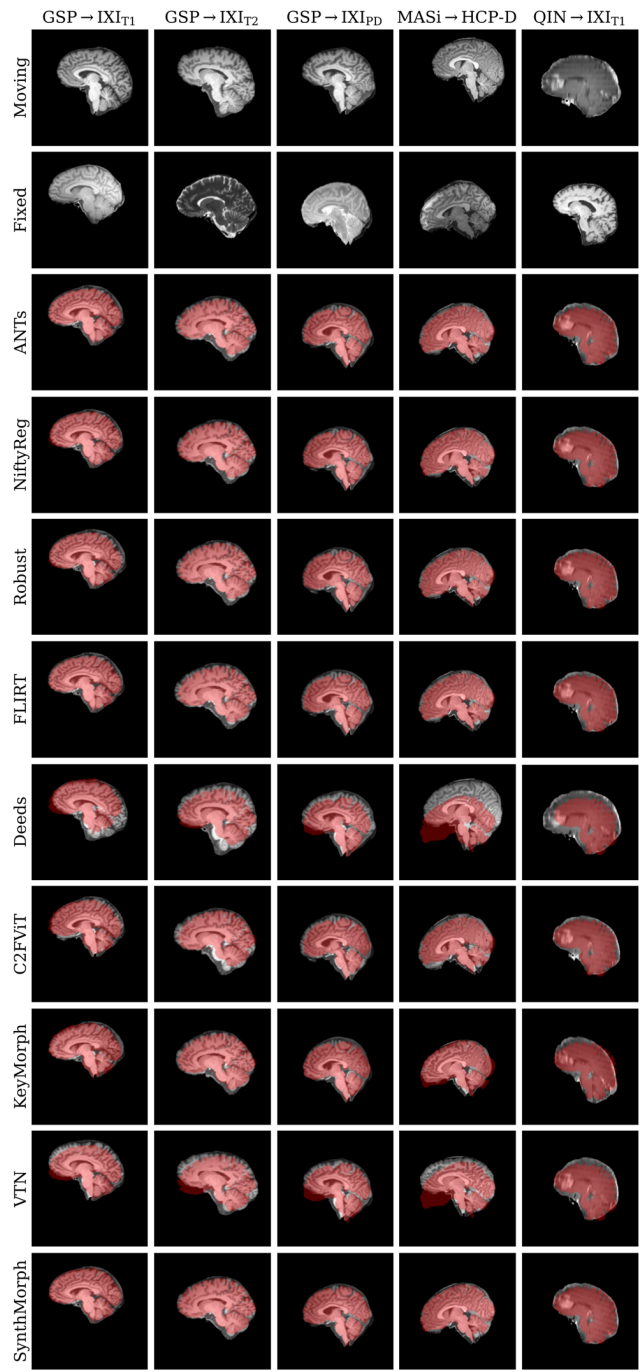

| GSP→IXI$_{T1}$ | GSP→IXI$_{T2}$ | GSP→IXI$_{PD}$ | MASi→HCP-D | QIN→IXI$_{T1}$ |

Moving · Fixed · ANTs · NiftyReg · Robust · FLIRT · Deeds · C2FViT · KeyMorph · VTN · SynthMorph

**Fig. 6.** Representative affine 3D registration examples showing the image moved by each method overlaid with the fixed brain mask (red). Each row is an example from a different dataset. Subscripts indicate MRI contrast.

ANTs accuracy even improves slightly, likely benefiting from the smoothing effect on the images. The classical baselines FLIRT and Robust are only mildly affected by thicker slices. While their Dice scores decrease more rapidly for $\Delta z \leq 8$, their accuracy reduces to 99% and about 98.5% at $\Delta z = 10$ mm. Deeds is noticeably more susceptible to resolution changes, decreasing to less than 95% at $\Delta z > 6.5$ mm.

Figure 10 compares the drop in median Dice overlap the affine methods undergo when presented with full-head as opposed to skull-stripped GSP→IXI$_{T1}$ images. Except for Deeds, brain-specific accuracy reduces substantially, by 3% in the case of NiftyReg and up to 8% for ANTs. Affine SynthMorph remains most robust: its Dice overlap changes by less than 0.05%. Deeds' accuracy increases but it still yields the lowest score for the testset.

Table 2 lists the registration time required by each affine method on a 2.2-GHz Intel Xeon Silver 4114 CPU using a single computational thread. The values shown reflect averages over $n = 10$ uni-modal runs. Classical runtimes range between 2 and 27 minutes, with Deeds being the fastest and Robust being the slowest, although we highlight that we substantially increased the number of Robust iterations. Complete single-threaded DL runtimes are about 1 minute, including model setup. However, inference only takes a few seconds and reduces to well under a second on an NVIDIA V100 GPU.

### 4.5. Experiment 2: joint registration

Motivated by the affine performance of SynthMorph, we complete the model with a hypernetwork-powered deformable module to achieve 3D joint affine-deformable registration (Fig. 4). Our focus is on building a complete and readily usable tool that generalizes across scan protocols without requiring preprocessing.

#### 4.5.1. Setup

First, we compare deformable registration using the held-out image pairs from separate subjects for each of the datasets of Table 1. The comparison employs skull-stripped images initialized with affine transforms estimated from skull-stripped data by NiftyReg, the most accurate baseline in Figure 5. We compare deformable SynthMorph performance to classical baselines and VTN, a joint DL baseline trained by the original authors—we seek to gauge the accuracy achievable with off-the-shelf algorithms for data unseen at training.

Second, we analyze the robustness of each tool to sub-optimal affine initialization. In order to cover realistic affine inaccuracies and assess the most likely and intended use case, we repeat the previous experiment, this time initializing each method with the affine transform obtained with the same method—that is, we test end-to-end joint registration with each tool. Similarly, we evaluate the importance of removing non-brain voxels from the input images. In this experiment, we initialize each method with affine transforms estimated by NiftyReg from skull-stripped data, and test deformable registration on a full-head version of the images.

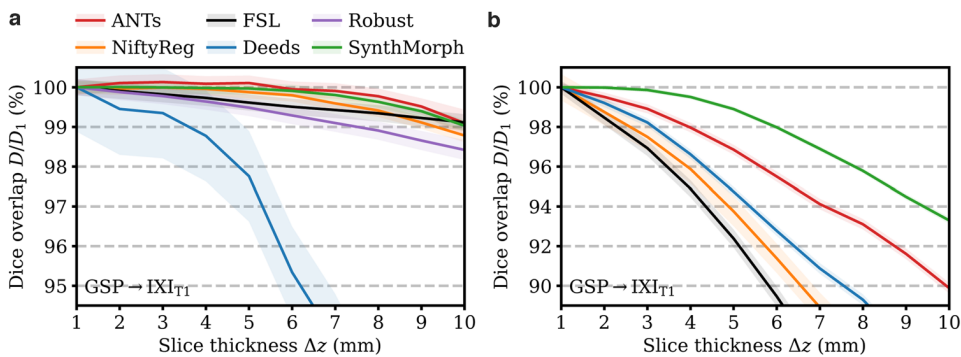

**Fig. 7.** Dependency of 3D (a) affine and (b) deformable registration accuracy on slice thickness. For comparability, we initialize all deformable tools with affine transforms estimated by NiftyReg. Each value indicates the mean over 100 skull-stripped pairs. Higher is better. Shaded areas indicate the standard error of the mean.

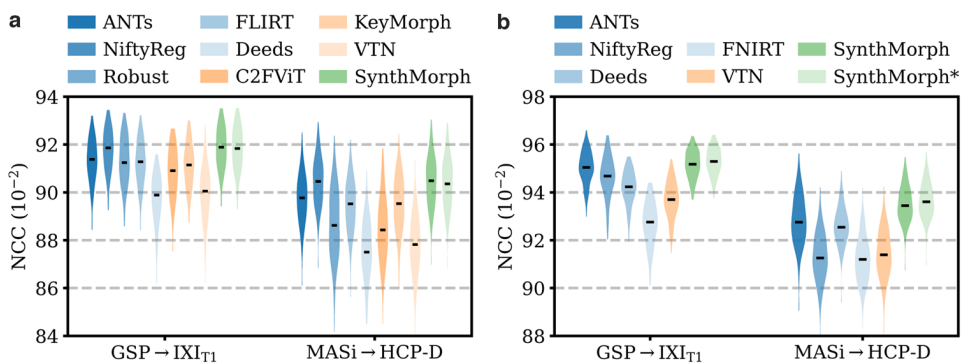

**Fig. 8.** Within-contrast 3D (a) affine and (b) deformable registration accuracy across skull-stripped cross-subject pairs in terms of brain-only NCC. We initialize all deformable tools with affine transforms estimated by NiftyReg. The asterisk indicates SynthMorph without skull-stripping. Higher is better.

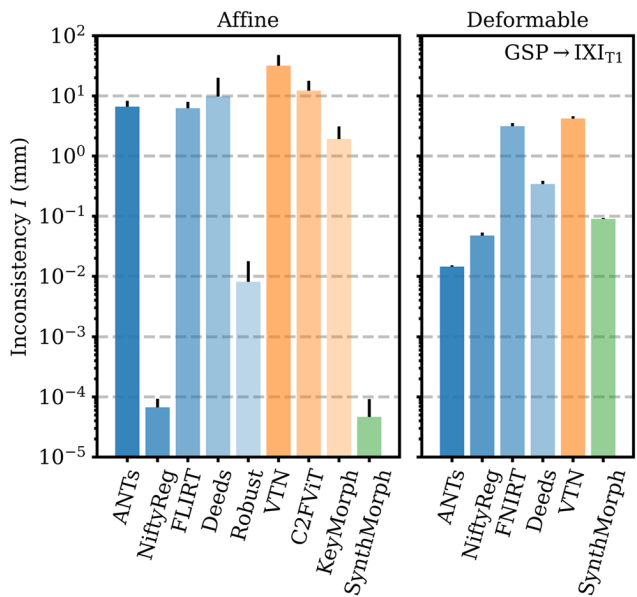

**Fig. 9.** Forward-backward inconsistency between transforms when reversing the order of input images. We compare the mean displacement per brain voxel upon subsequent application of both transforms. Lower values are better.

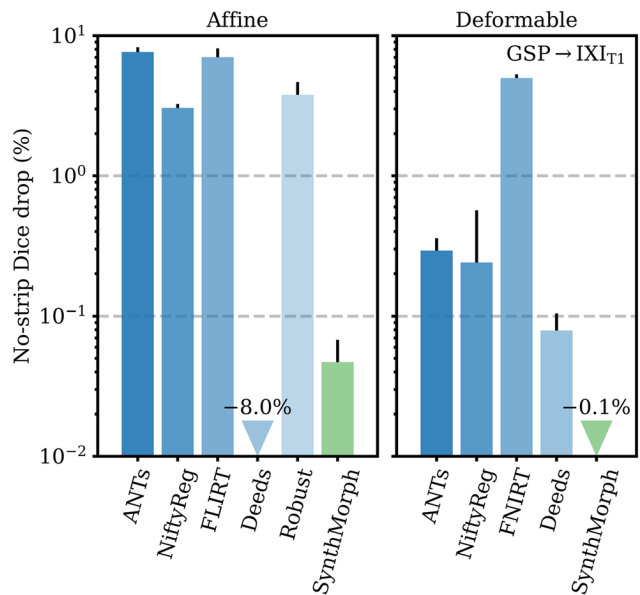

**Fig. 10.** Relative reduction in brain-specific accuracy when registering full-head as opposed to skull-stripped images. Lower values are better. Although affine Deeds is the only method whose Dice overlap increases, it ranks as the least accurate on the GSP→IXI$_{T1}$ testset. Error bars show the standard error of the mean.

Third, we analyze the effect of reducing the through-plane resolution $\Delta z$ on SynthMorph performance compared to classical baselines, following the steps outlined in Section 4.4. In this experiment, we initialize each method with affine transforms estimated by NiftyReg from skull-stripped images, such that the comparison solely reflects deformable registration accuracy.

Fourth, we analyze warp-field regularity and registration accuracy over dataset GSP→IXI$_{T1}$ as a function of the regularization weight $\lambda$. We also compare the symmetry of each method with regard to reversing the order of the input images.

### 4.5.2. Results

Figure 11 shows typical deformable registration examples for each method, and Figure 12 compares registration accuracy across testsets in terms of mean Dice overlap $D$ over the 21 largest anatomical structures (large-21), 10 fine-grained structures (small-10) not optimized at training, and image similarity measured with MSE-MIND. Supplementary Figures S1–S5 show deformable registration accuracy across individual brain structures.

Although SynthMorph trains with synthetic images only, it achieves the highest large-21 score for every skull-stripped testset. For all cross-contrast pairings and the pediatric testset, SynthMorph leads by at least 2 Dice points compared to the highest baseline score (MASi→HCP-D, $p < 10^{-23}$ for paired two-sided $t$-test) and often much more. Across these testsets, SynthMorph performance remains largely invariant, whereas the other methods except Deeds struggle. Crucially, the distribution of SynthMorph scores for isotropic data is substantially narrower than the baseline scores, indicating the absence of gross inaccuracies such as pairs with $D < 65$ that several baselines yield across all isotropic contrast pairings. On the clinical testset QIN→IXI$_{T1}$, SynthMorph surpasses the baselines by at least $\Delta D = 4$. For GSP→IXI$_{T1}$, it outperforms the best classical baseline ANTs by 1 Dice point ($p < 10^{-21}$).

Across the T1w testsets, FNIRT outperforms NiftyReg by several Dice points and also ANTs for MASi → HCP-D pairs. Surprisingly, FNIRT beats NiftyReg's NMI implementation for GSP→IXI$_{T2}$, even though FNIRT's cost function targets within-contrast registration. The most robust baseline is Deeds, which ranks third at adult T1w registration. Its performance reduces the least for the cross-contrast and clinical testsets, where it achieves the highest Dice overlap after SynthMorph.

The joint DL baseline VTN yields relatively low accuracy across all testsets. This was expected for the cross-contrast pairings, since the model was trained with T1w data, confirming the data dependency introduced with standard training. However, VTN lags behind the worst-performing classical baseline for GSP→IXI$_{T1}$ data, NiftyReg, too ($\Delta D = 2.1$, $p < 3 \times 10^{-7}$), likely due to domain shift as in the affine case.

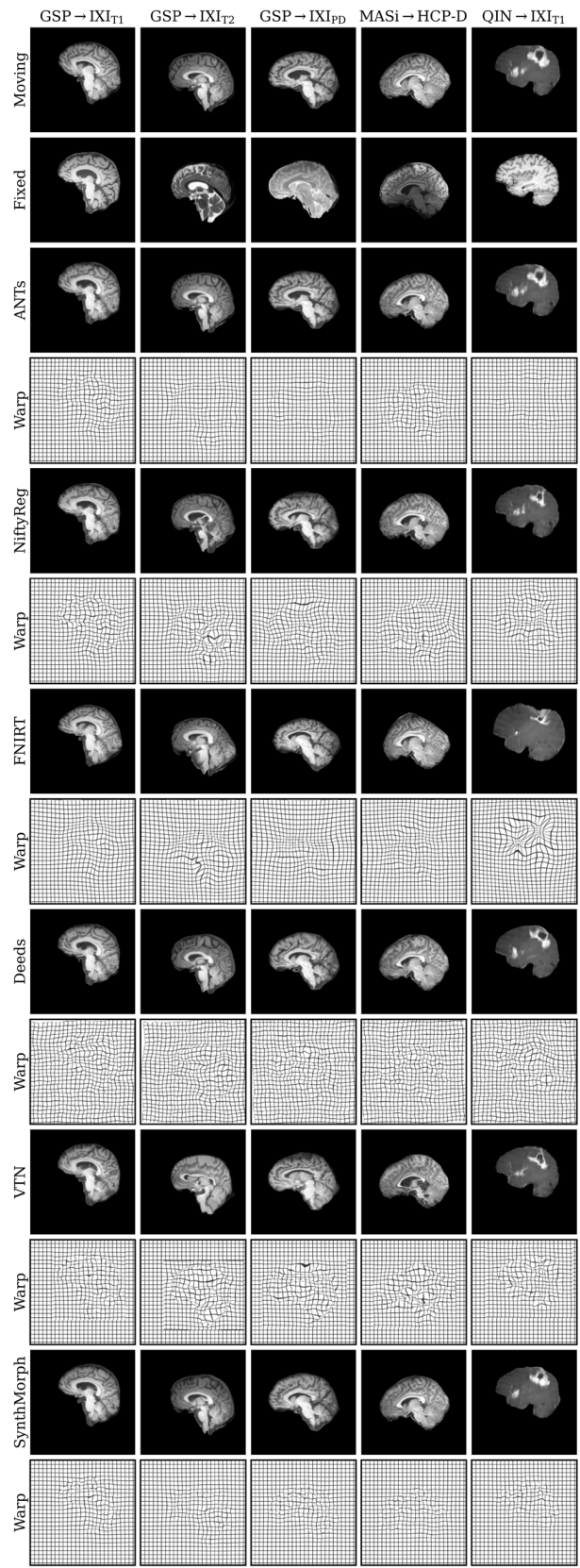

**Fig. 11.** Deformable 3D registration examples comparing the moved image $m \circ \phi$ and the deformation field $\phi$ across methods. Each row is an example from a different dataset. For comparability, we initialize all methods with NiftyReg's affine registration.

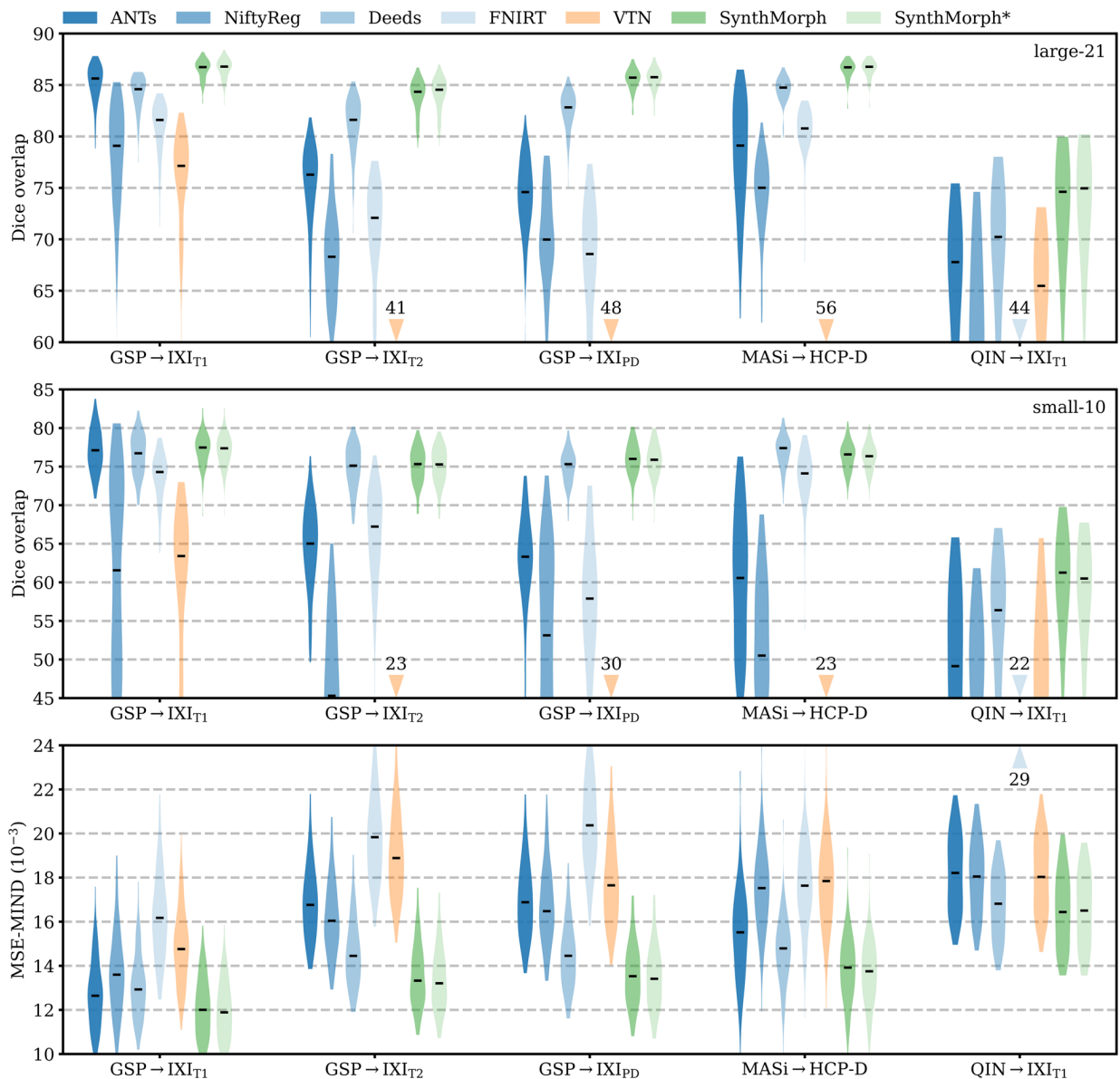

**Fig. 12.** Deformable 3D registration accuracy as mean Dice scores over the 21 largest brain regions (large-21), 10 fine-grained structures not optimized at SynthMorph training (small-10), and image similarity. Each violin shows the distribution across the skull-stripped cross-subject pairs from Table 1. For comparability, we initialize all deformable tools with affine transforms estimated by NiftyReg. The asterisk indicates SynthMorph performance without skull-stripping. Downward arrows show median scores outside the plotted range. Higher Dice and lower MSE-MIND are better.

Considering the fine-grained small-10 brain structures held out at training, SynthMorph consistently matches or exceeds the best performing method, except for MASi→HCP-D, where Deeds leads by $\Delta D = 0.6$ ($p = 10^{-4}$). On the clinical testset, SynthMorph leads by at least $\Delta D > 4.5$ ($p < 10^{-8}$). Interestingly, SynthMorph outperforms all baselines across testsets in terms of MSE-MIND ($p < 10^{-4}$) and

NCC for same-contrast registration (Fig. 8b, $p < 10^{-3}$), although it is the only method not optimizing or trained with an image-based loss.

Figure 13 shows the relative change in large-21 Dice for each tool when run end-to-end compared to affine initialization with NiftyReg. SynthMorph's drop in performance is 0.05% or less across all datasets. For GSP →IXI$_{T1}$, classical-baseline accuracy decreases by no more than 0.3%. Across the other datasets, the classical methods generally cannot make up for the discrepancy between their own and NiftyReg's affine transform: accuracy drops by up to 5.2%, whereas SynthMorph remains robust. The performance of VTN reduces by at least 8.3% across testsets and often much more, highlighting the detrimental effect an inaccurate affine transform can have on the subsequent deformable step.

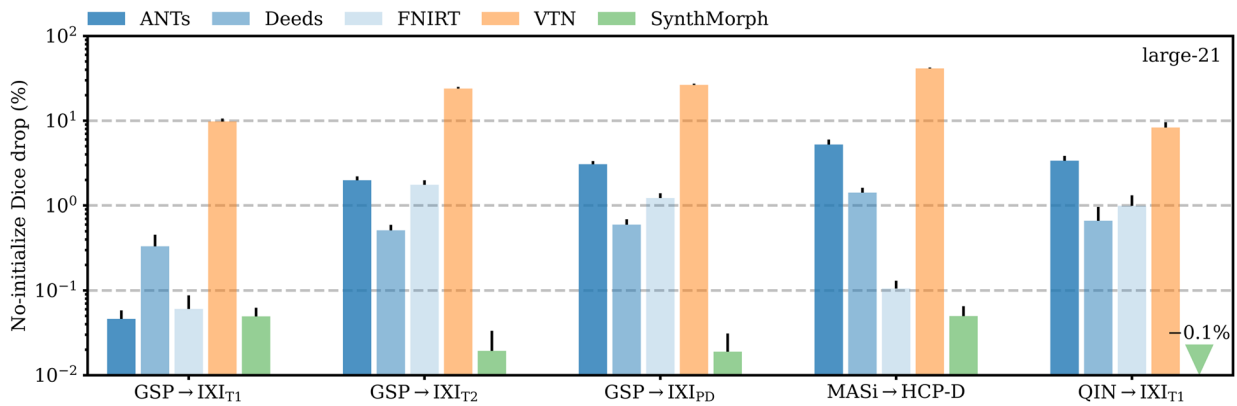

**Fig. 13.** Mean decrease in Dice scores for end-to-end joint registration relative to affine initialization with NiftyReg. Except for adult T1w registration pairs, the classical tools in blue generally cannot compensate for the discrepancy between their own and NiftyReg's affine transform, indicating that inaccurate affine initialization can have a detrimental effect on subsequent deformable registration.

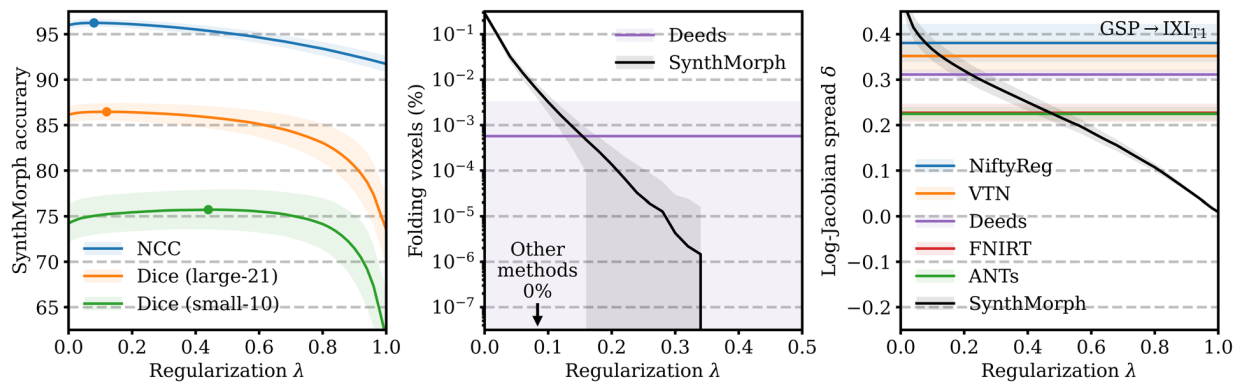

**Fig. 14.** Regularization analysis of SynthMorph registration accuracy, the proportion of folding voxels with a negative Jacobian determinant, and the spread of the distribution of absolute log-Jacobian determinants as a function of the regularization weight $\lambda$. The dots indicate maximum accuracy. For the other metrics, lower is better.

Figure 10 shows the importance of skull-stripping for deformable registration accuracy. Generally, deformable accuracy suffers less than affine registration when switching to full-head images, as the algorithms deform image regions independently. SynthMorph remains most robust to the change in preprocessing; its large-21 Dice overlap increases by 0.1%. With a drop of 0.08%, Deeds is similarly robust. In contrast, FNIRT's performance is most affected, reducing by 5%—a decline of the same order as for most affine methods.

Figure 14 analyzes SynthMorph warp smoothness. As expected, image-based NCC and large-21 Dice accuracy peak for weak regularization of $\lambda < 0.2$. In contrast, overlap of the small-10 regions not optimized at training benefits from smoother warps, with an optimum at $\lambda = 0.45$. The fields predicted by SynthMorph achieve the lowest log-Jacobian spread across all baselines for $\lambda > 0.45$. Similarly, the proportion of folding brain voxels decreases with higher $\lambda$ and drops to 0% for $\lambda > 0.33$ (10 integration steps). Deeds yields $6 \times 10^{-4}$% folding brain voxels, whereas the other baselines achieve 0%. For realistic warp fields with characteristics that match or exceed the tested baselines, we conduct all comparisons in this study with a default weight $\lambda = 0.5$. We highlight that $\lambda$ is an input to SynthMorph, enabling users to choose the optimal regularization strength for their specific data without retraining.

Deformable registration with SynthMorph is highly symmetric (Fig. 9), with a mean forward-backward inconsistency of only $I = 0.09$ mm per brain voxel that closely follows ANTs (0.01 mm) and NiftyReg (0.05 mm). In contrast, the remaining methods are substantially more inconsistent, with $I = 0.34$ mm for Deeds or more.

Figure 7b assesses the dependency of registration performance on slice thickness $\Delta z$. Similar to the affine case, deformable accuracy decreases for thicker slices, albeit faster. SynthMorph performs most robustly. Its accuracy remains unchanged up to $\Delta z \leq 3$ mm and reduces only to 95% at $\Delta z = 8.5$ mm. ANTs is the most robust classical method, but its accuracy drops considerably faster than SynthMorph. FLIRT and NiftyReg are most affected at reduced resolution, performing at less than 95% accuracy for $\Delta z \geq 4$ mm and $\Delta z \geq 4.5$ mm, respectively.

Deformable registration often requires substantially more time than affine registration (Table 2). On the GPU, SynthMorph takes less than 8 seconds per image pair for registration, IO, and resampling. One-time model setup requires about 1 minute, after which the user could register any number of image pairs without reinitializing the model. SynthMorph requires about 16 GB of GPU memory for affine and 24 GB for deformable registration. On the CPU, the fastest classical method Deeds requires only about 6 minutes in single-threaded mode, whereas ANTs takes almost 5 hours. While VTN's joint runtime is 1 minute, SynthMorph needs about 15 minutes for deformable registration on a single thread.

## 5. DISCUSSION

We present an easy-to-use DL tool for end-to-end affine and deformable brain registration. SynthMorph achieves robust performance across acquisition characteristics such as imaging contrast, resolution, and pathology, enabling accurate registration for brain scans without preprocessing. The SynthMorph strategy alleviates the dependency on acquired training data by generating widely variable images from anatomical label maps—and there is no need for label maps at registration time.

### 5.1. Anatomy-specific registration

Accurate registration of the specific anatomy of interest requires ignoring or down-weighting the contribution of irrelevant image content to the optimization metric. SynthMorph *learns* what anatomy is pertinent to the task, as we optimize the overlap of select labels of interest only. It is likely that the model learns an implicit segmentation of the image, in the sense that it focuses on deforming the anatomy of interest, warping the remainder of the image only to satisfy regularization constraints. In contrast, many existing classical and DL methods cannot distinguish between relevant and irrelevant image features, and thus have to rely on explicit segmentation to remove distracting content prior to registration, such as skull-stripping (Eskildsen et al., 2012; Hoopes, Mora, et al., 2022; Iglesias et al., 2011; Smith, 2002).

Pathology missing from the training labels does not necessarily hamper overall registration accuracy, as the experiments with scans from patients with glioblastoma show. In fact, SynthMorph outperforms all deformable baselines tested on these data. However, we do not expect these missing structures to be mapped with high accuracy, in particular if the structure is absent in one of the test images—this is no different from the behavior of methods optimizing image similarity.

### 5.2. Baseline performance

Networks trained with the SynthMorph strategy do not have access to the MRI contrasts of the testsets nor, in fact, to any MRI data at all. Yet SynthMorph matches or outperforms classical and DL-baseline performance across the real-world datasets tested, while being substantially faster than the classical methods. For deformable registration, the fastest classical method Deeds requires 6 minutes, while SynthMorph takes about 1 minute for one-time model setup and just under 8 seconds for each subsequent registration. This speed-up may be particularly useful for processing large datasets like ABCD, enabling end-to-end registration of hundreds of image pairs per hour—the time that some established tools like ANTs require for a single registration.

The DL baselines tested have runtimes comparable to SynthMorph. Combining them with skull-stripping would generally be a viable option for fast brain-specific registration: brain extraction with a tool like SynthStrip only takes about 30 seconds. However, we are not aware of any existing DL tool that would enable deformable registration of unseen data with adjustable regularization strength *without* retraining. While the DL baselines break down for contrast pairings unobserved at training, they also cannot match the accuracy of classical tools for the T1w contrast they were trained with, likely due to domain shift.

In contrast, SynthMorph performance is relatively unaffected by changes in imaging contrast, resolution, or subject population. These results demonstrate that the SynthMorph strategy produces powerful networks that can register new image types unseen at training. We emphasize that our focus is on leveraging the training strategy to build a robust and accurate registration tool. It is possible that other architectures, such as the trained DL baselines tested in this work, perform equally well when trained using our strategy. Specifically, novel Bayesian similarity learning methods (Grzech et al., 2022; Su & Yang, 2023) and frameworks that jointly optimize the affine and deformable components emerged since the initial submission of this work (Chang et al., 2023; Meng et al., 2023; Qiu et al., 2023; L. Zhao et al., 2023).

Although Robust down-weights the contribution of image regions that cannot be mapped with the linear transformation model of choice, its accuracy dropped by several points for data without skull-stripping. The poor performance in cross-contrast registration may be due to the experimental nature of its robust-entropy cost function. We initially experimented with the recommended NMI metric, but registration failed for a number of cases as Robust produced non-invertible matrix transforms, and we hoped that the robust metrics would deliver accurate results in the presence of non-brain image content—which the NMI metric cannot ignore during optimization.

### 5.3. Challenges with retraining baselines

Retraining DL baselines to improve performance for specific user data involves substantial practical challenges. For example, users have to reimplement the architecture and training setup from scratch if code is not available. If code is available, the user may be unfamiliar with the specific programming language or machine-learning library, and building on the original authors' implementation typically requires setting up an often complex development environment with matching package versions. In our experience, not all authors make this version information readily available, such that users may have to resort to trial and error. Additionally, the user's hardware might not be on par with the authors'. If a network exhausts the memory of the user's GPU, avoiding prohibitively long training times on the CPU necessitates reducing model capacity, which can affect performance. We emphasize that because SynthMorph registers new images without retraining, it does not require a GPU. On the CPU, SynthMorph runtimes still compare favorably to classical methods (Table 2).

In principle, users could retrain DL methods despite the above-mentioned challenges. However, in practice the burden is usually sufficiently large that users of these technologies will turn to methods that distribute pretrained models. For this reason, we specifically compare DL baselines trained by the respective authors, to gauge the performance attainable without retraining. While our previous work (Hoffmann, Billot, et al., 2021) demonstrated the feasibility of training registration networks within the synthesis strategy and, in fact, without any acquired data at all, the original model predicted implausibly under-regularized warps, and changing the regularization strength required retraining. In contrast, the toolbox version provides fast, domain-robust, symmetric, invertible, general-purpose DL registration, enabling users to choose the optimal regularization strength for their specific data—without retraining. We hope that the broad applicability of SynthMorph may help alleviate the historically limited reusability of DL methods.

EasyReg (Iglesias, 2023) is a recent DL registration method developed concurrently with SynthMorph. Both methods leverage the same synthesis strategy (Hoffmann et al., 2022) and thus do not require retraining. They differ in that EasyReg fits an affine transform to hard segmentation maps and estimates transforms to MNI space internally (Fonov et al., 2009), whereas SynthMorph includes an affine registration network and estimates pair-wise transforms directly. In addition, SynthMorph enables the user to control the warp smoothness at test time.

### 5.4. Joint registration

The joint baseline comparison highlights that deformable algorithms cannot always fully compensate for real-world inaccuracies in affine initialization. Generally, the median Dice overlap drops by a few percent when we initialize each tool with affine transforms estimated by the same package instead of NiftyReg, the most accurate affine baseline we tested. This experiment demonstrates the importance of affine registration for joint accuracy—choosing affine and deformable algorithms from the same package is likely the most common use case.

In Section 4.4, the affine subnetwork of the 10-cascade VTN model consistently ranks among the least accurate methods even for the T1w image type it trained with. We highlight that the authors of VTN do not independently tune or compare the affine component to baselines and instead focus on joint affine-deformable accuracy (S. Zhao, Dong, et al., 2019; S. Zhao, Lau, et al., 2019). While the VTN publication presents the affine cascade as an Encoder architecture ($C = 1$, Section A.1) terminating with an FC layer (S. Zhao, Lau, et al., 2019), the public implementation omits the FC layer. Some of our experiments with this architecture indicated that the FC layer is critical to competitive performance.

### 5.5. Limitations

While SynthMorph often achieves state-of-the-art performance, we also discuss several limitations. First, the large-21 evaluation of registration accuracy uses the same anatomical labels whose overlap SynthMorph training optimizes. Although the analyses also compare the small-10 labels not optimized at training, MSE-MIND, and NCC, we consider only one label for the left and another for the right cortex, limiting the evaluation predominantly to subcortical alignment.

Second, some applications require fewer DOF than the full affine matrix that SynthMorph estimates. For example, the bulk motion in brain MRI and its mitigation through pulse-sequence adjustments are constrained to 6 DOF accounting for translation and rotation (Gallichan et al., 2016; Singh et al., 2024; Tisdall et al., 2012; White et al., 2010). Although the SynthMorph utility includes a model for rigid alignment trained with scaling and shear (Appendix B) removed from matrix $\hat{t}$ of Equation (7), the evaluation focuses on affine registration.

Third, considering voxel data alone, the SynthMorph rotational range is limited as the model only sees registration pairs rotated by angles below $|r_i| = 180°$ about any axis $i$, resulting from the rotational offset between any two input label maps combined with spatial augmentation (Appendix Table A2), because the affine model did not converge with augmentation across the full range $r_i \in [-180°, 180°]$. However, the registration problem reduces to an effective 90° range when considering the orientation information stored in medical image headers. Ignoring headers, the rotational ranges measured across OASIS and ABCD do not exceed $|r_i| \leq 43.1°$ (Appendix Fig. A4).

Fourth, we train SynthMorph as a general tool for cross-subject registration, and the evaluation on clinical data is limited to 50 glioblastoma patients.

In addition, accuracy for specialized applications such as tumor tracking will likely trail behind dedicated models. However, for tumor-specific training, our learning framework could add synthesized pathology to label maps from healthy subjects. For example, an extended synthesis may simulate the mass effect by applying deformations measured in healthy-pathologic image pairs (Hogea et al., 2007) and overlaying the deformed label map with a synthetic tumor label (Zhou et al., 2023) to subsequently generate a distinct image intensity.

### 5.6. Future work

We plan to expand our work in several ways. First, we will provide a trained 6-DOF model for rigid registration, as many applications require translations and rotations only, and the most accurate rigid transform does not necessarily correspond to the translation and rotation encoded in the most accurate affine transform.

Second, we will employ the proposed strategy and affine architecture to train specialized models for within-subject registration for navigator-based motion correction of neuroimaging with MRI (Gallichan et al., 2016; Hoffmann et al., 2016; Tisdall et al., 2012; White et al., 2010). These models need to be efficient for real-time use but do not have to be invariant to MRI contrast or resolution when employed to track head-pose changes between navigators acquired with a fixed protocol. However, the brain-specific registration made possible by SynthMorph will improve motion-tracking and thus correction accuracy in the presence of jaw movement (Hoffmann et al., 2020).

Third, another application that can dramatically benefit from anatomy-specific registration is fetal neuroimaging, where the fetal brain is surrounded by amniotic fluid and maternal tissue. We plan to tackle registration of the fetal brain, which is challenging, partly due to its small size, and which currently relies on brain extraction prior to registration to remove confounding image content (Billot, Moyer, et al., 2023; Gaudfernau et al., 2021; Hoffmann, Abaci Turk, et al., 2021; Puonti et al., 2016).

## 6. CONCLUSION

We present an easy-to-use DL tool for fast, symmetric, diffeomorphic—and thus invertible—end-to-end registration of images without preprocessing. Our study demonstrates the feasibility of training accurate affine and joint registration networks that generalize to image types unseen at training, outperforming established baselines across a landscape of image contrasts and resolutions. In a rigorous analysis approximating the diversity of real-world data, we find that our networks achieve invariance to protocol-specific image characteristics by leveraging a strategy that synthesizes widely variable training images from label maps.

Optimizing the spatial overlap of select anatomical labels enables anatomy-specific registration without the need for segmentation that removes distracting content from the input images. We believe this independence from complex preprocessing has great promise for time-critical applications, such as real-time motion correction of MRI. Importantly, SynthMorph is a widely applicable learning strategy for anatomy-aware and acquisition-agnostic registration of any anatomy with any network architecture, as long as label maps are available for training—there is no need for these at registration time.

### DATA AND CODE AVAILABILITY

A stand-alone SynthMorph utility and the source code are available at https://w3id.org/synthmorph. We also distribute SynthMorph as part of the open-source FreeSurfer package at https://freesurfer.net. The experiments presented in this study retrospectively analyze public datasets whose original sources we indicate in Section 4.1.

### AUTHOR CONTRIBUTIONS

Malte Hoffmann worked on conceptualization, data curation, formal analysis, funding acquisition, investigation, methodology, project administration, software, and writing (original draft, review, editing). Andrew Hoopes assisted with data curation, methodology, resources, software, validation, and writing (review, editing). Douglas N. Greve helped with data curation, funding acquisition, resources, validation, and writing (review, editing). Bruce Fischl worked on conceptualization, funding acquisition, investigation, methodology, resources, software, supervision, and writing (review, editing). Adrian V. Dalca contributed to conceptualization, methodology, resources, software, supervision, and writing (review, editing).

## ETHICS

We retrospectively re-analyze public datasets only. The Mass General Brigham Internal Review Board (IRB) approved the FSM study and the infant study. The Committee on Clinical Investigation at Boston Children's Hospital approved the infant study. Retrospective analysis of de-identified open-access data did not require ethical approval. All subjects gave written informed consent.

## DECLARATION OF COMPETING INTEREST

Bruce Fischl has financial interests in CorticoMetrics, a company whose medical pursuits focus on brain imaging and measurement technologies. Malte Hoffmann and Bruce Fischl received salary support from GE Health-Care. Massachusetts General Hospital and Mass General Brigham manage this interest in accordance with their conflict of interest policies. The authors have no other interests that could have inappropriately influenced the work reported in this paper.

## ACKNOWLEDGEMENTS

The authors thank Lilla Zöllei and OASIS Cross-Sectional for sharing data (PIs D. Marcus, R. Buckner, J. Csernansky, J. Morris, NIH awards P50 AG05681, P01 AG03991, P01 AG026276, R01 AG021910, P20 MH071616, U24 R021382). We are grateful for funding from the National Institute of Biomedical Imaging and Bioengineering (P41 EB015896, P41 EB030006, R01 EB019956, R01 EB023281, R01 EB033773, R21 EB018907), the National Institute of Child Health and Human Development (R00 HD101553), the National Institute on Aging (R01 AG016495, R01 AG070988, R21 AG082082, R56 AG064027), the National Institute of Mental Health (RF1 MH121885, RF1 MH123195), the National Institute of Neurological Disorders and Stroke (R01 NS070963, R01 NS083534, R01 NS105820, U01 NS132181, UM1 NS132358), and the National Center for Research Resources (S10 RR019307, S10 RR023043, S10 RR023401). We acknowledge additional support from the BRAIN Initiative Cell Census Network (U01 MH117023) and the NIH Blueprint for Neuroscience Research (U01 MH093765), part of the multi-institutional Human Connectome Project. The study also benefited from computational hardware provided by the Massachusetts Life Sciences Center.

## SUPPLEMENTARY MATERIALS

Supplementary material for this article is available with the online version here: https://doi.org/10.1162/imag_a_00197.

## A. AFFINE NETWORK ANALYSIS

### A.1. Affine architectures

We analyze and compare three competing network architectures (Appendix Fig. A1) that represent state-of-the art methods (Balakrishnan et al., 2019; De Vos et al., 2019; Moyer et al., 2021; Shen et al., 2019; Yu et al., 2021; Zhu et al., 2021): Decoder from Section 3.3.1 and the following Encoder and Decomposer architectures.

#### A.1.1. Parameter encoder

We build on networks combining a convolutional encoder with an FC layer (Shen et al., 2019; Zhu et al., 2021) whose $N(N+1)$ output units we interpret as parameters for translation, rotation, scale, and shear. We refer to a cascade of $C$ such subnetworks $h_i$, with $i \in \{1, 2, ..., C\}$, as "Encoder". Each $h_i$ outputs a matrix constructed from the affine parameters as shown in Appendix B, to incrementally update the total transform. We obtain transform $T_i$ by matrix multiplication after invoking subnetwork $h_i$,

$$T_i = h_1(m_0, f)\, h_2(m_1, f) \cdots h_i(m_{i-1}, f) \qquad (A1)$$

where $m_i = m \circ T_i$ is the moving image transformed by $T_i$, and $T_0 = I_N$ is the identity matrix. As the subnetworks $h_i$ are architecturally identical, weight sharing is possible, and we evaluate versions of the model with and without weights shared across cascades.

For balanced gradient steps, we complete each subnetwork with a layer applying a learnable rescaling weight to each affine parameter before matrix construction.

#### A.1.2. Warp decomposer

We propose another architecture building on deformable registration models (Balakrishnan et al., 2019; De Vos et al., 2019). "Decomposer" estimates a dense deformation field $\phi_\theta$ with corresponding non-negative voxel weights $\kappa_\theta$ that we decompose into the affine output transform $T_\theta = h_\theta(m, f)$ and a (discarded) residual component $\delta_\theta$, that is, $\phi_\theta = \delta_\theta \circ T_\theta$. The voxel weights $\kappa_\theta$ enable the network $h_\theta$ to focus the decomposition on the anatomy of interest. Both $(\phi_\theta, \kappa_\theta)$ are outputs of a single fully convolutional network and thus benefit from weight sharing. We decompose $\phi_\theta$ in a WLS sense over the spatial domain $\Omega$ of $f$, using the definition of $t$ from Equation (1) as the submatrix of $T$ that excludes the last row:

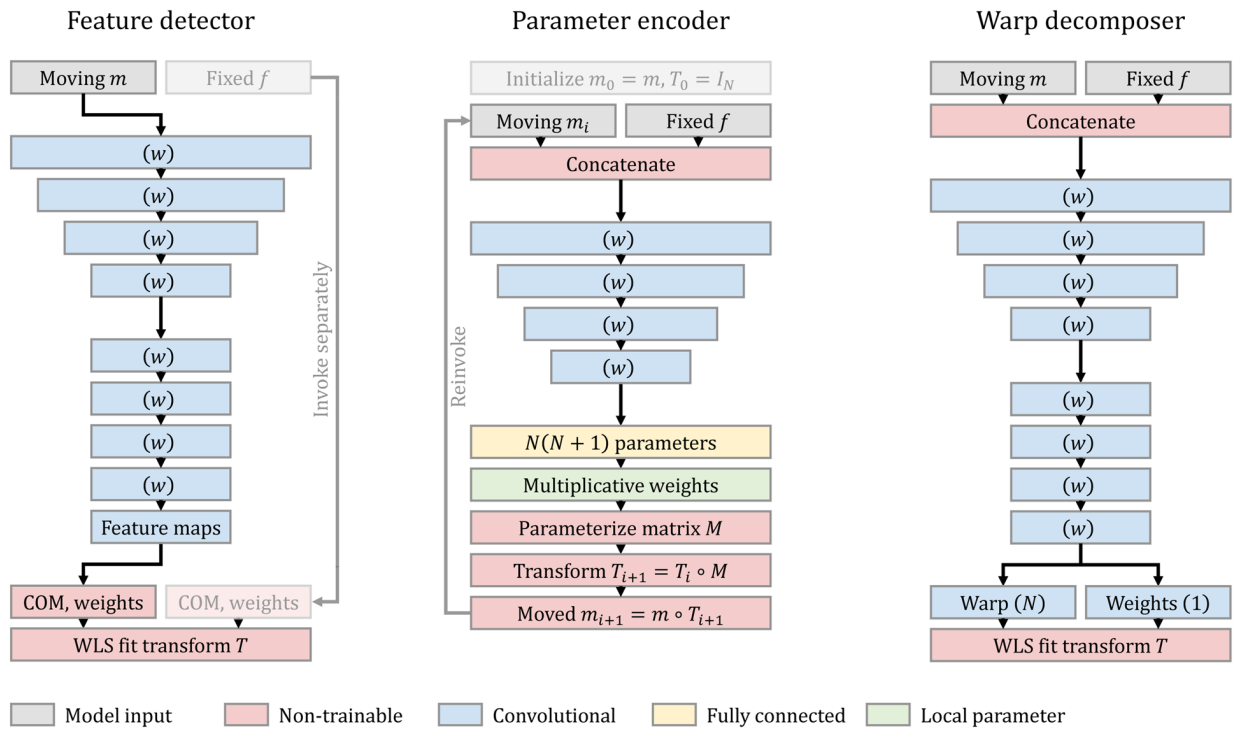

**Appendix Fig. A1.** Affine architectures. Detector outputs ReLU-activated feature maps for a single image. We compute their centers of mass (COM) and weights separately for $m$ and $f$, to fit a transform $T$ that aligns these point sets. A recurrent Encoder estimates refinements to the current transform $T_i$ from moved image $m_i = m \circ T_i$ and fixed image $f$. Decomposer predicts a one-shot displacement field (no activation) with corresponding voxel weights (ReLU), that we decompose in a weighted least-squares (WLS) sense to estimate affine transform $T$. Parentheses specify filter numbers. We LeakyReLU-activate the output of unnamed convolutional blocks (param. $\alpha = 0.2$). Stacked convolutional blocks of decreasing size indicate subsampling by a factor of 2 via max pooling following each activation.

$$\hat{t}_\theta = \arg\min_t \sum_{x \in \Omega} \kappa_\theta(x) \| \phi_\theta(x)^\top - (x^\top \quad 1)t t^\top \|^2, \quad (A2)$$

where $t^\top$ is the matrix transpose of $t$. Denoting $W = \mathrm{diag}(\kappa_\theta)$, and by $X$ and $y$ the matrices whose corresponding rows are $(x^\top \; 1)$ and $\phi_\theta(x)^\top$ for each $x \in \Omega$, respectively, Equation (5) yields the closed-form WLS solution as in Section 3.3.1.

### A.1.3. Implementation and training

Encoder predicts rotation parameters in degrees. This parameterization ensures varying rotation angles has an effect of similar magnitude as translations in millimeters, at the scale of the brain, which helps networks converge faster in our experiments. We initialize the rescaling weights of Encoder to 1 for translations and rotations, and to 0.05 for scaling and shear, which we find favorable to faster convergence. Appendix B includes details.

Training optimizes an unsupervised NCC loss between the moved image $m \circ h_\theta(m, f)$ and the fixed image $f$. All models train for a single strip with a batch size of 2 (Section 3.3.5). To avoid non-invertible matrices $M = X^\top WX$ at the start of training, we pretrain Decomposer for 500 iterations, temporarily replacing the output transform with the field $T_\theta = \phi_\theta \odot \kappa_\theta$, where $\kappa_\theta$ are the voxel weights predicted by the network (Section A.1.2), and $\odot$ denotes voxel-wise multiplication.

### A.2. Data

For architecture analysis, we use T1w images with isotropic 1-mm resolution from adult participants aged 40–75 years from the UK Biobank (UKBB) study (Alfaro-Almagro et al., 2018; Miller et al., 2016; Sudlow et al., 2015). We conform images and derive label maps as in Section 4.1, extracting mid-sagittal slices from corresponding 3D images and label maps.

### A.3. Experiment

Assuming a network capacity of ~250 k learnable parameters, we explore the relative strengths and weaknesses of each affine architecture. We conduct the experiment in a 2D-registration context, which reduces the computational burden to consider numerous model configurations.

### A.3.1. Setup

We train networks drawing $\{m, f\}$ from a set of 5000 images, and test registration on a validation set of 100 distinct cross-subject pairs that does not overlap with the training set. To keep network capacities comparable,

**Appendix Table A1.** Network capacity for model comparison.

| Architecture | Config. | $w$ | Capacity | Dev. (%) |
|---|---|---|---|---|
| Encoder | $C = 2^0$ | 72 | 252 k | +0.8 |
| Encoder | $C = 2^1$ | 45 | 250 k | 0.0 |
| Encoder | $C = 2^2$ | 27 | 247 k | +1.2 |
| Encoder | $C = 2^3$ | 16 | 255 k | +2.0 |
| Encoder | $C = 2^4$ | 9 | 260 k | +4.0 |
| Decomposer | $n = 0$ | 63 | 253 k | +1.2 |
| Decomposer | $n = 1$ | 59 | 254 k | +1.6 |
| Decomposer | $n = 2$ | 55 | 248 k | −0.8 |
| Decomposer | $n = 3$ | 52 | 246 k | −1.6 |
| Detector | $k = 2^3$ | 62 | 248 k | −0.8 |
| Detector | $k = 2^4$ | 62 | 252 k | +0.8 |
| Detector | $k = 2^5$ | 61 | 253 k | +1.2 |
| Detector | $k = 2^6$ | 58 | 246 k | −1.6 |
| Encoder | $C = 2^0$ | 110 | 498 k | −0.4 |
| Decomposer | $n = 0$ | 89 | 504 k | +0.8 |
| Detector | $k = 2^5$ | 87 | 503 k | +0.6 |

Each model configuration uses a different network width $w$ held constant across its convolutional layers to reach a target capacity of 250 k or 500 k parameters, up to a small deviation.

each model uses a different width $w$, held constant across its convolutional layers (Appendix Table A1). Training uses only the affine augmentation indicated in Appendix Table A2.

First, we test if Encoder benefits from weight sharing. We train separate models with $C \in \{1, 2, 4, 8, 16\}$ subnetworks that either share or use separate weights.

Second, we compare Decomposer variants that fit $T$ in an OLS sense, using weights $\forall x \in \Omega$, $\kappa_\theta(x) = 1$, or in a WLS sense. For both, we assess the impact of the resolution $\varrho$ of the field $\phi_\theta$ relative to $f$ on performance, by upsampling by a factor of 2 after each of the first $n \in \{0, 1, 2, 3\}$ convolutional decoder blocks, using skip connections where possible, such that $\varrho(n) = 1 / 2^{4-n}$. A resolution of $\varrho = 1 / 4$ corresponds to 25% of the resolution of $f$, that is, 25% of the original image dimensions.

Third, we analyze OLS and WLS variants of Detector predicting $k \in \{8, 16, 32, 64\}$ feature maps to compute the corresponding $(a_i, p_i)$ and $(b_i, q_i)$ for $i \in \{1, 2, \ldots, k\}$.

Finally, we select one configuration per architecture and analyze its performance across a range of transformation magnitudes. We investigate how models adapt to larger transforms, by fine-tuning trained weights to twice the affine augmentation amplitudes of Appendix Table A2 until convergence, and we repeat the experiment with doubled capacity. The test considers copies of the test set, applying random affine transforms of maximum strength $\gamma \in [0, 2]$ relative to the augmentation range of Appendix Table A2. For example, at a given $\gamma$, we uniformly sample a rotation angle $r \sim \mathcal{U}(-\gamma\alpha, \gamma\alpha)$ for each of the 200 moving and fixed images, where $\alpha = 45°$, and similarly for the other 6 degrees of freedom (DOF, Appendix B).

### A.3.2. Results

Appendix Figure A2 compares registration accuracy for the NCC-trained models in terms of Dice overlap. Encoder achieves the highest accuracy, surpassing the best Detector configuration by up to 0.4 and the best Decomposer by up to 1 Dice point. Using more subnetworks improves Encoder performance, albeit with diminishing returns after $C = 4$ and at the cost of substantially longer training times that roughly scale with $C$. The local rescaling weights of subnetwork $h_1$ converge to values around 1 for translations and rotations and around 0.01 for scaling and shear. The values tend to decrease for subsequent $h_i$ ($i > 1$), most noticeably the translational weights. For $h_C$, the translational weights converge to roughly 50% of those of $h_1$, suggesting that the first subnetworks perform the bulk of the alignment, whereas the subsequent $h_i$ refine it by smaller amounts. Although keeping subnetwork weights separate might also enable each $h_i$ to specialize in increasingly fine adjustments to the final transform, in practice we observe no benefit in distributing capacity over the subnetworks compared to weight sharing.

Decomposer shows a clear trend toward lower output resolutions $\varrho$ improving accuracy. Although decomposing the field $\phi_\theta$ in a WLS sense boosts performance by 0.6–1.3 points over OLS, the model still lags behind the other architectures while requiring 2–3 times more training iterations to converge. There is little difference across numbers $k$ of output feature maps, and choosing WLS over OLS results in a minor increase in accuracy.

Appendix Figure A3 shows network robustness across a range of maximum transform strengths $\gamma$, where we compare Encoder with $C = 4$ subnetworks sharing weights to the WLS variants of Decomposer without upsampling and Detector with $k = 32$ output channels, to balance performance and efficiency. Detector proves most robust to large transforms, remaining unaffected up to $\gamma \approx 1.2$, that is, shifts and rotations up to 36 mm and 54° for each axis, respectively, and scale and shear up to 0.12. In contrast, accuracy declines substantially for Encoder and Decomposer after $\gamma \approx 0.8$, corresponding to maximum transforms of 24 mm and 36° (blue). Doubling the affine augmentation extends Encoder and Decomposer robustness to $\gamma \approx 1.2$ but comes at the cost of a drop of 1 and 2 Dice points for all $\gamma < 1.2$, respectively (orange). Decomposer performance is capacity-bound, as doubling the number of parameters restores ~50% of the drop in accuracy, whereas increasing capacity does not improve Encoder accuracy for $\gamma < 1.2$ (green). Detector optimally benefits from the doubled affine augmentation, which helps the network perform

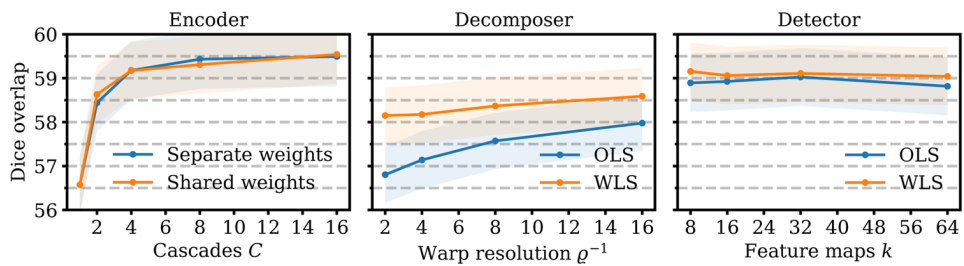

**Appendix Fig. A2.** Network analysis. We assess Encoder with different numbers of subnetworks $C$. We also analyze Decomposer and Detector variants using ordinary (OLS) or weighted least squares (WLS), varying the warp resolution $\varrho$ and number of output feature maps $k$, respectively. A value of $\varrho^{-1} = 4$, for example, means the warp $\phi_\theta$ has resolution and dimensions of only 25% compared to the input images. Dice scores are averages over 100 UKBB cross-subject 2D pairs. Shaded areas indicate the standard error of the mean.

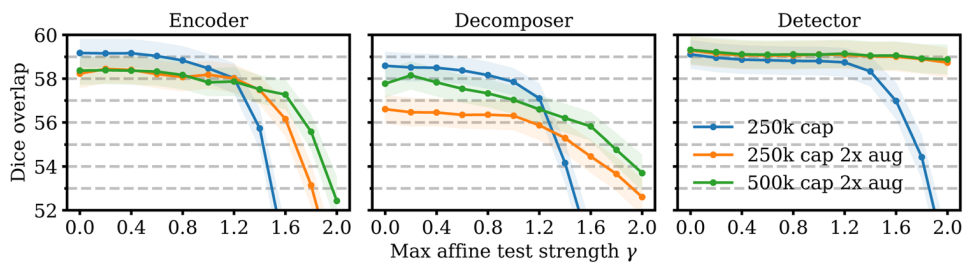

**Appendix Fig. A3.** Network robustness across affine transform strengths $\gamma$ relative to the range of Appendix Table A2. At a given $\gamma$, we resample each image, drawing affine parameters from uniform distributions modulated by $\gamma$, such as angle $r \sim \mathcal{U}(-\gamma\alpha, \gamma\alpha)$ with $\alpha = 45°$. We test models trained with doubled augmentation (aug) and capacity (cap), comparing $C = 4$ Encoders sharing weights, the WLS Decomposer without upsampling, and a $k = 32$ Detector. Dice scores are averages over 100 UKBB cross-subject 2D pairs. Shaded areas indicate the standard error of the mean.

robustly across the test range (orange). Doubling capacity has no effect (green).

In conclusion, the marginal lead of Encoder only manifests for small transforms and at $C = 16$ cascades. The 16 interpolation steps of this variant render it intractably inefficient for 3D applications.

In contrast, Detector performs with high accuracy across transformation strengths, making it a more suitable architecture for a general registration tool.

### A.4. Discussion

The network analysis shows that Encoder is an excellent architecture if the expected transforms are small, especially at a number of cascades $C \geq 4$. For medium to large transforms, Encoder accuracy suffers. While the experiments indicate that the reduction in accuracy can be mitigated by simultaneously optimizing a separate loss for each cascade, doing so substantially increases training times compared to the other architectures. Another drawback of Encoder is the image-size dependence introduced by the FC layer. Detector is a more flexible alternative that remains robust for medium to large transforms. While the results of a 2D analysis may not generalize fully to 3D registration, prior work confirms the robustness of Detector across large transforms in 3D (Yu et al., 2021).

Vision transformers (Dosovitskiy et al., 2020) are another popular approach to overcoming the local receptive field of convolutions with small kernel sizes, querying information across distributed image patches. However, in practice, the sophisticated architecture is often unnecessary for many computer-vision tasks (Pinto et al., 2022): while simple small-kernel U-Nets generally perform well as their multi-resolution convolutions effectively widen the receptive field (Z. Liu et al., 2022), increasing the kernel size can boost the performance of convolutional networks beyond that achieved by vision transformers across multiple tasks (Ding et al., 2022; S. Liu et al., 2022).

### B. AFFINE PARAMETERIZATION

Let $f$ be a fixed $N$D image of side lengths $d_i$, where $i \in \{1, \dots, N\}$ indexes the right-handed axes of the spatial image domain $\Omega$. This work uses zero-centered index voxel coordinates $x \in \Omega$. That is,

$$\Omega = \prod_{i=1}^{N} \{-\Delta d_i, \ 1 - \Delta d_i, \ \dots, d_i - 1 - \Delta d_i\} \tag{A3}$$

with $\Delta d_i = (d_i - 1) / 2$, placing the center of rotations at the center of $f$. Let $T : \Omega \to \mathbb{R}^N$ be the affine coordinate transform of Equation (1), which maps a moving image $m$ onto the domain of $f$. We parameterize $T$ as

$$T = VRZE, \tag{A4}$$

where $V$, $R$, $Z$, $E$ are matrices describing translation, rotation, scaling, and shear, respectively. Denoting by $v_i$ the translation and by $z_i$ the scaling parameter along axis $i$, we define

$$V = \left( \begin{array}{ccc|c} & & & v_1 \\ & I_N & & \vdots \\ & & & v_N \\ \hline 0 & \cdots & 0 & 1 \end{array} \right), \tag{A5}$$

and

$$Z = \left( \begin{array}{cccc} z_1 & & & \\ & \ddots & & \\ & & z_N & \\ & & & 1 \end{array} \right). \tag{A6}$$

For rotations and shear, we distinguish between the 2D and 3D case. Let $r_i$ be the angle of rotation about axis $i$, where the direction of rotation follows the right-hand rule. We abbreviate $c_i = \cos(r_i)$ and $s_i = \sin(r_i)$.

### B.1. Two-dimensional case

In 2D, we apply the rotation angle $r = r_3$ and shear $e$ using matrices

$$R = \left( \begin{array}{ccc} c_3 & -s_3 & 0 \\ s_3 & c_3 & 0 \\ 0 & 0 & 1 \end{array} \right) \text{ and } E = \left( \begin{array}{ccc} 1 & e & 0 \\ 0 & 1 & 0 \\ 0 & 0 & 1 \end{array} \right). \tag{A7}$$

### B.2. Three-dimensional case

We consider intrinsic 3D rotations represented as the matrix product $R = R_1 R_2 R_3$, where

$$R_1 = \left( \begin{array}{cccc} 1 & 0 & 0 & 0 \\ 0 & c_1 & -s_1 & 0 \\ 0 & s_1 & c_1 & 0 \\ 0 & 0 & 0 & 1 \end{array} \right), \tag{A8}$$

$$R_2 = \left( \begin{array}{cccc} c_2 & 0 & s_2 & 0 \\ 0 & 1 & 0 & 0 \\ -s_2 & 0 & c_2 & 0 \\ 0 & 0 & 0 & 1 \end{array} \right), \tag{A9}$$

$$R_3 = \begin{pmatrix} c_3 & -s_3 & 0 & 0 \\ s_3 & c_3 & 0 & 0 \\ 0 & 0 & 1 & 0 \\ 0 & 0 & 0 & 1 \end{pmatrix}, \tag{A10}$$

and we apply the shears $\{e_i\}$ using the parameterization:

$$E = \begin{pmatrix} 1 & e_1 & e_2 & 0 \\ 0 & 1 & e_3 & 0 \\ 0 & 0 & 1 & 0 \\ 0 & 0 & 0 & 1 \end{pmatrix}. \tag{A11}$$

### B.3. Transforming coordinates

With the notation introduced in Equation (1), we transform the coordinates of an $N$D point

$$x = \begin{pmatrix} x_1 & x_2 & \dots & x_N \end{pmatrix}^\top \in \Omega \tag{A12}$$

as $x' = Ax + v$, or, using a single matrix product,

$$\left( \frac{x'}{1} \right) = T \left( \frac{x}{1} \right). \tag{A13}$$

## C. GENERATION HYPERPARAMETERS

Appendix Table A2 lists the generation hyperparameter ranges that SynthMorph training uses for label-map augmentation and image synthesis.

## D. TRANSFORM ANALYSIS

In this supplementary experiment, we analyze the range of typical transforms a registration tool may have to cope with. We register 1000 distinct and randomly pooled subject pairs from OASIS and another 1000 pairs from ABCD. The estimated transformation matrix $T$ decomposes into the translation, rotation, scaling, and shearing parameters defined in Appendix B.

Appendix Figure A4 presents the range of absolute transformation parameters measured within OASIS and ABCD, along any axis $i \in \{1, 2, 3\}$ of the common space introduced in Section 4.1. Within OASIS, the mean translations and rotations are $\overline{|v_i|} = (8.5 \pm 9.9)$ mm and $\overline{|r_i|} = (5.6 \pm 6.0)°$, respectively ($\pm$ standard deviation, SD). The average scaling and shearing parameters are $\overline{|z_i - 1|} = (5.0 \pm 3.8)\%$ and $\overline{|e_i|} = (3.3 \pm 3.0)\%$, respectively. While the bulk of the transforms is small, a subset of the OASIS subjects are far apart, leading to large total ranges of translation $|v_i| \leq 61.1$ mm, rotation $|r_i| \leq 43.1°$, scaling $|z_i - 1| \leq 22.6\%$, and shear $|e_i| \leq 22.1\%$. Transforms within ABCD follow a similar distribution.

**Appendix Table A2.** Uniform hyperparameter sampling ranges $[a, b]$ for synthesizing training images from source segmentation maps.

| Hyperparameter | Unit | $a$ | $b$ |
|---|---|---|---|
| Translation | mm | −30 | 30 |
| Rotation | ° | −45 | 45 |
| Scaling | % | 90 | 110 |
| Shear | % | 90 | 110 |
| Warp sampling SD | mm | 0 | 2 |
| Warp blurring FWHM | mm | 8 | 32 |
| Label intensity mean | a.u. | 0 | 1 |
| Noise intensity SD | % | 10 | 20 |
| Image blurring FWHM | mm | 0 | 8 |
| Bias field sampling SD | % | 0 | 10 |
| Bias field blurring FWHM | mm | 48 | 64 |
| FOV cropping | % | 0 | 20 |
| Downsampling factor | % | 1 | 8 |
| Gamma exponent | – | 0.5 | 1.5 |

We abbreviate standard deviation (SD), full width at half maximum (FWHM), and field of view (FOV).

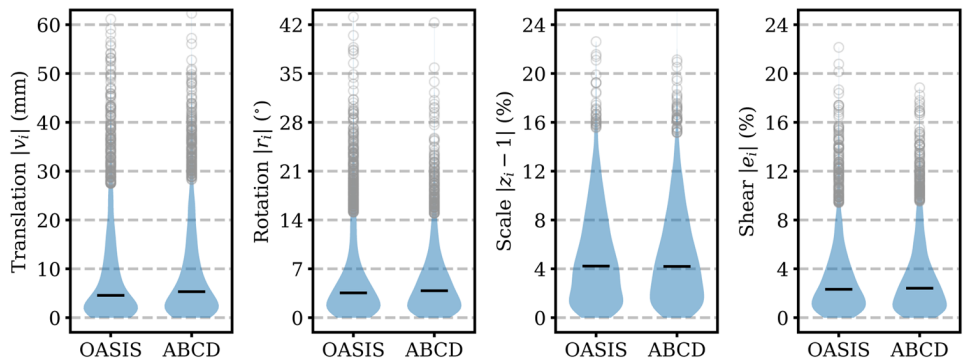

**Appendix Fig. A4.** Absolute affine transformation range across $n = 1000$ registration pairs randomly selected from either OASIS or ABCD. Each panel pools parameters relative to all axes $i \in \{1, 2, 3\}$ of 3D space. Black bars indicate median values. Circles represent parameters farther than 1.5 inter-quartile ranges from the median.

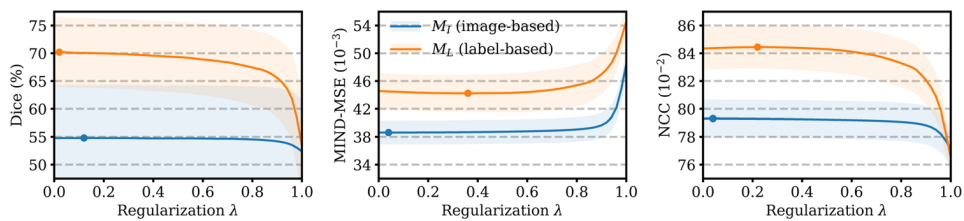

**Appendix Fig. A5.** Deformable registration accuracy at test across 100 T1-weighted cross-subject UKBB pairs after training with label-based versus image-based loss terms. Apart from the optimized similarity loss term, the trained models are identical. The dots indicate optimum accuracy: higher is better, except for MIND-MSE. Shaded areas indicate the standard error of the mean.

Therefore, we choose to augment input label maps at training with affine transforms drawn from the ranges of Appendix Table A2, ensuring that SynthMorph covers the transformation parameters measured across public datasets.

## E. LOSS COMPARISON

In another supplementary experiment, we explore the benefit of optimizing a loss on one-hot encoded label maps compared to optimizing a multi-modal image-similarity loss. We train two identical deformable hyper-models on synthetic images: one model optimizing MSE on brain labels (Section 3.3.3), the other optimizing the modality-independent image-similarity loss MIND-MSE over the brain (Section 4.3). For both models, we complete the similarity term with the regularization term of Equation (10), and we align all training label maps and evaluation images to a common affine space using NiftyReg.

Assessing brain-registration accuracy across 100 cross-subject pairs from the UKBB (Section A.2), we find that training with the label-based loss leads to better Dice scores at test, while the image-based loss leads to better MIND-MSE scores at test (Appendix Fig. A5). However, label-based training substantially outperforms training with MIND-MSE in terms of the image-similarity metric NCC—which, unfortunately, we cannot straightforwardly optimize within the synthesis-based training strategy as it does not perform well for image pairs with different contrasts (Hoffmann et al., 2022).

The image-based loss term may still be sensitive to some of the contrast and artifact differences between the fixed and moving training images that do not represent differences in anatomy, whereas the label-based loss is independent of these intensity differences by construction.